\documentclass[journal,onecolumn]{IEEEtran}
\makeatletter
\def\ps@headings{%
\def\@oddhead{\mbox{}\scriptsize\rightmark \hfil \thepage}%
\def\@evenhead{\scriptsize\thepage \hfil \leftmark\mbox{}}%
\def\@oddfoot{}%
\def\@evenfoot{}}
\makeatother 

\def\P{\mathbb{P}}
\def\conv{*}

\def\eps{\varepsilon}

\def\et{\textit{et al.}}

\def\conv{*}

\def\eps{\varepsilon}

\def\R{{\mathbb{R}}}

\usepackage{amsmath}
\usepackage{colordvi}
\usepackage{amsfonts}
\usepackage{graphicx}
\usepackage{epsfig}
\usepackage{color}
\usepackage{url}

\newtheorem{theorem}{Theorem}

 {\begin{list}{}%
         {\setlength{\leftmargin}{#1}}%
         \item[]%
 }
 {\end{list}}

\begin{document}

\title{Towards a System Theoretic Approach to Wireless Network Capacity in Finite Time and
Space}

\author{Florin Ciucu, Ramin Khalili, Yuming Jiang, Liu Yang, Yong Cui

\thanks{Florin Ciucu and Ramin Khalili are with Telekom Innovation Laboratories / TU Berlin, Germany, e-mails: \{florin,ramin\}@net.t-labs.tu-berlin.de.}
\thanks{Yuming Jiang is with the Department of Computer Science, Norwegian University of Science and Technology, Norway, e-mail: ymjiang@ieee.org}
\thanks{Liu Yang and Yong Cui are with the Department of Computer Science, Tsinghua University, e-mails: 09232022@bjtu.edu.cn and cuiyong@tsinghua.edu.cn}
}

\maketitle

\begin{abstract}
In asymptotic regimes, both in time and space (network size), the
derivation of network capacity results is grossly simplified by
brushing aside queueing behavior in non-Jackson networks. This
simplifying double-limit model, however, lends itself to
conservative numerical results in finite regimes. To properly
account for queueing behavior beyond a simple calculus based on
average rates, we advocate a system theoretic methodology for the
capacity problem in finite time and space regimes. This
methodology also accounts for spatial correlations arising in
networks with CSMA/CA scheduling and it delivers rigorous
closed-form capacity results in terms of probability
distributions. Unlike numerous existing asymptotic results,
subject to anecdotal practical concerns, our transient one can be
used in practical settings: for example, to compute the time
scales at which multi-hop routing is more advantageous than
single-hop routing.
\end{abstract}


\section{Introduction}
The fields of information theory and communication networks have
been for long evolving in isolation of each other, in what is
referred to as an unconsummated union (Ephremides and
Hajek~\cite{EphremidesH98}). A fundamental cause is that
information theory ignores data burstiness and delay, which are
typical manifestations in communication (queueing) networks. While
data burstiness has little relevance for the point to point
channel since the receiver is practically oblivious to when data
is received, it should however be properly accounted for the
multiaccess channel since the time required for all the
transmitters to appear smoothed-out may be far longer than
tolerable delays (Gallager~\cite{Gallager85}). Furthermore, when
accounting for random arrivals, the burstiness is even more
critical for a tandem of point to point channels, and networks in
general.

A groundbreaking work at the intersection of the two fields is a
set of results obtained by Gupta and Kumar~\cite{Gupta00}. Under
some simplifications at the network layers (e.g., no multi-user
coding schemes, or ideal assumptions on power-control, routing,
and scheduling), the authors derived network capacity results as
asymptotic scaling laws on the maximal data rates which can be
reliably sustained in multi-hop wireless networks. The elegance
and importance of these results have been very inspirational,
especially within the networking community.

The results in~\cite{Gupta00}, and of related work, are based on
technical arguments involving a \textit{double-limit model}. The
outer limit is explicitly taken in the number of nodes $n$---
capturing an \textit{infinite-space} model---in order to guarantee
certain structural properties in random networks with high
probability. The inner limit is implicitly taken in
time---capturing an \textit{infinite-time} model---and which
enables a simple calculus based on average rates to derive upper
and lower bounds on network capacity. The double-limit model can
be regarded as being reminiscent of information theory and
relating itself to the \textit{infinite-space} model employed in
the analysis of the multiaccess channel (i.e., infinitely many
sources are assumed to coexist)~\cite{Gallager85}.

The key advantage of the technical arguments from~\cite{Gupta00}
is that all nodes appear as smoothed-out at the data link layer
and the network capacity analysis is drastically simplified;
indeed, by solely reasoning in terms of average rates (first
moments), the difficult problem of accounting for burstiness
(e.g., higher moments) in non-Jackson queueing networks is
avoided. While such a calculus is mathematically justified in
asymptotic regimes, its implications in finite regimes have been
largely evaded so far; by `finite regime' we mean both finite time
and finite number of nodes.

To shed light in the direction of computing the network capacity
in finite regimes, this paper makes three contributions:
\begin{enumerate}
\item[C1.] It discusses the fundamental limitations in finite
regimes of the double-limit argument from~\cite{Gupta00}.
Concretely, the direct reproduction of asymptotic techniques in
finite regimes does not capture a non-negligible factor for both
the upper and lower capacity bounds, which thus justifies the
anecdotal impracticality of numerous asymptotic results. These
findings motivate the need for alternative analytical techniques
to compute the network capacity in finite time and space regimes,
beyond the convenient but simplistic averages-based calculus.

\item[C2.] It advocates a \textit{system theoretic approach} to
the \textit{transient} network capacity problem at the per-flow
level. The crucial advantage of this approach is that it
conveniently deals with inherent queueing behavior at downstream
nodes. Moreover, it also copes with spatial correlations arising
in networks with CSMA/CA scheduling, in the sense that no
artificial assumptions (e.g., statistical independence) are
necessary.

\item[C3.] It illustrates the applicability of finite time and
space capacity results to decide when multi-hop routing is
theoretically more advantageous than single-hop routing. The paper
shows the time scale at which the lower bound (on throughput
capacity) for the former is greater than the upper bound for the
latter, and thus indicates the \textit{time scales} at which
multi-hop is the most advantageous.
\end{enumerate}

From a technical point of view, the main ideas of the advocated
methodology mentioned in Item C2 lies on a subtle analogy between
single-hop links and linear time invariant (LTI) systems, by
constructing impulse-responses to entirely characterize successful
transmissions over single-hop links. The impulse-responses are
closed under a convolution operator, which conveniently accounts
for queueing behavior at downstream nodes. These ideas have been
recently explored by Ciucu~\et~\cite{CiHoHu10,CiucuISIT11,CiSc13}
for the particular Aloha protocol. This paper \textit{generalizes}
these prior works by formulating a \textit{unified}
system-theoretic framework which additionally captures two more
MAC protocols: centralized scheduling and especially the
challenging CSMA/CA.

An advantage of the proposed framework is that it yields capacity
results in terms of probability distributions, and thus all the
moments, including average rates or variances, are readily
available. Moreover, the capacity results are directly obtained at
the per-flow level. Such a per-flow analysis can provide
information about the fairness of routing and scheduling
algorithms and hence could be useful in protocol design. As
multiple paths are available between source-destination pair, one
can use this information to provide route optimization and load
balancing in the network. The concrete practical application
addressed in the paper was described in Item C3.

The rest of the paper is organized as follows. In
Section~\ref{sec:dtc} we discuss the limitations of the technical
arguments from \cite{Gupta00} based on a double-limit model in
finite time and space regimes. In Section~\ref{sec:worksol} we
introduce the advocated system theoretic methodology to derive
capacity results in finite regimes. In Section~\ref{sec:mac} we
show how to fit three MAC protocols in this methodology, and
further synthesize the derivation of transient capacity results
including the case of dynamic networks. Section~\ref{sec:shmh}
presents the multi-hop vs. single-hop practical application.
Section~\ref{sec:numerics} gives additional numerical results and
Section~\ref{sec:discussion} concludes the paper.

\section{On the Limitations of The Double-Limit Model}\label{sec:dtc}
Consider the random network model from Gupta and
Kumar~\cite{Gupta00} in which $n$ nodes are uniformly placed on a
disk/square of area one. For each node in the network, a random
destination is chosen such that there are $n$ source-destination
pairs. We consider the Protocol Model from~\cite{Gupta00}, which
defines successful transmission in terms of Euclidean distances.

The capacity problem is about finding the maximum value of
$\lambda(n)$, i.e., the rate of each transmission, guaranteeing
network stability. Computing upper and lower bounds on
$\lambda(n)$ is based on a simple calculus involving the
end-to-end (e2e) transmissions' average rates at the relay nodes,
which are implicitly subject to a time limit. For some arbitrary
e2e transmission $i$, let $\tilde{\lambda}_{i,j}(n)$ denote the
incoming average rate at some arbitrary node $j$, i.e.,
\begin{equation*}
\tilde{\lambda}_{i,j}(n)=\limsup_{t\rightarrow\infty}\frac{A_{i,j}(t)}{t}~,
\end{equation*}
where $A_{i,j}(t)$ denotes the cumulative arrival process and $t$
denotes time. In general it holds that
$\tilde{\lambda}_{i,j}(n)\leq\lambda(n)$, whereas an exact
relationship depends on many factors such as routing, scheduling,
or the network stability; such factors may also lend themselves to
conceivable scenarios in which the `$\lim$' does not exist, whence
the more technical `$\limsup$' definition above.

With the new notation, one can rephrase the network capacity
problem as finding the maximal rate $\lambda(n)$ such that
\begin{equation*}
\lambda(n)=\tilde{\lambda}_{i,j}(n)~\forall i,j~.
\end{equation*}

The main result from~\cite{Gupta00} is that
\begin{equation}
\lambda(n)=\Theta\left(\frac{1}{\sqrt{n\log{n}}}\right)~.\label{eq:gkcap}
\end{equation}
Here, the underlying space limit in $n$ guarantees useful
structural properties in the considered random network with high
probability (e.g., e2e connectivity or bounds on the number of
transit transmissions at some node).

Capacity results such as the one from Eq.~(\ref{eq:gkcap}) are
based on a double-limit model, in which the outer (space) limit is
in $n$ whereas the inner (time) limit is in $t$. En passant, it is
interesting to observe that the limits are not interchangeable;
indeed, note that by letting the outer limit in $t$, the rates at
downstream relay nodes tend to zero (e.g., when $n>t$). More
interestingly, a single-limit model can be considered by suitably
letting $t$ as a function of $n$. Depending on structural network
properties, the rate at which $t$ should increase could be as
large as
\begin{equation*}
t=\omega\left(n^2\right)~.
\end{equation*}
This is necessary, for instance, in the following scenario: $n$
nodes numbered as $\left\{1,2,\dots,n\right\}$ are placed around a
circle, every node $i$ transmits to the counter-clockwise neighbor
$(i+n-2)\%n+1$ along the clockwise path
$i\%n+1,(i+1)\%n+1,\dots,(i+n-2)\%n+1$, and all transmissions
interfere with each other (`$\%$' is the modulo operation). Under
a perfect scheduling, we remark that at most $k$ delivered packets
from all $n$ e2e transmissions could be guaranteed in $kn^2$
slots, for any $k$, whence the $\omega(n^2)$ lower bound.

Next we discuss the numerical implications of the double-limit
model on existing bounds on $\lambda(n)$; in such a double-limit
setting, we assume a single limit in $n$ and a suitable (implicit)
limit in t, e.g., $t=\omega(n^2)$.

\subsection{A Calculus for $\lambda(n)$}\label{eq:calc}
We revisit the key ideas from~\cite{Gupta00} to compute upper and
lower bounds on $\lambda(n)$. We argue in particular that both
(asymptotic) bounds do not capture a non-negligible
multiplicative/fractional factor, which means that the bounds can
be quite loose in finite regimes; Subsection~\ref{sec:simus}
provides related numerical results.

\subsubsection{Upper bounds}
The underlying idea is based on the condition
\begin{equation}
n\lambda(n) h\leq x~,\label{eq:gkar}
\end{equation}
where $h$ is a lower bound on the number of average hops, whereas
$x$ is an upper bound on the number of simultaneous and successful
active nodes (see p.~402, $2^{\textrm{nd}}$ column,
$1^{\textrm{st}}$ equation from~\cite{Gupta00}). The left-hand
side (LHS) is thus a lower bound on how much information
\textit{must} be transmitted, assuming a rate $\lambda(n)$ for
each source, whereas the right-hand side (RHS) is an upper bound
on how much information \textit{can} be transmitted (note that
both LHS and RHS are asymptotic \textit{rates}, i.e., time
averages of some stochastic processes). For the random network
model from~\cite{Gupta00},
$h=\Theta\left(\sqrt{\frac{n}{\log{n}}}\right)$ and
$x=\Theta\left(\frac{n}{\log{n}}\right)$; the two asymptotic
expressions are sufficient to guarantee structural properties in
the random network model from~\cite{Gupta00} with high
probability.

The upper-bound argument from Eq.~(\ref{eq:gkar}) was extensively
adapted in the network capacity literature: see, e.g., Eq.~(2) in
Li~\et~\cite{Li01} for unicast capacity in static ad hoc networks,
Eqs.~(18,20) in Mergen and Tong~\cite{Mergen05} for unicast
capacity in networks with regular structure, Eq.~(26) in~Neely and
Modiano~\cite{NeelyM05a} for unicast capacity in some mobile
networks, Eq.~(2) in Shakkottai~\et~\cite{ShakkottaiLS10} for
multicast capacity, etc.), and even in some much earlier papers by
Kleinrock and Silvester~(see Eq.~(18) in \cite{KleinrockSilv78}
for unicast capacity in uniform random networks and
Eqs.~(12,25,29) in~\cite{Silvester83} for unicast capacity in
networks with regular structure, both employing the Aloha
protocol).

Let us now discuss the validity of the upper-bound argument in
more restrictive space/time models. In a finite-space (fixed $n$)
infinite-time model, the argument also holds subject to further
conditions: e2e paths must exist for all source-destination pairs
and (non-asymptotic) expressions for $h$ and $x$ are known. Under
the same structural conditions, the upper-bound continues to hold
in a finite time/space model (fixed $n$ and time span $T$) by
properly interpreting rates over finite time intervals.

What is interesting to observe in the finite regime is that
Eq.~(\ref{eq:gkar}) can be tightened as
\begin{equation}n\lambda(n)h\leq(1-f(n,T))x~,\label{eq:ub2}
\end{equation}
where the factor $f(n,T)$ denotes the average \textit{percentage}
of the number of empty buffers. Indeed, over a finite time span
$T$, the average number of simultaneous and successful
transmissions decays due to transient burstiness effects, in
particular due to the existence of empty buffers at the very nodes
scheduled to transmit.

A last more minor observation concerns whether the bound from
Eq.~(\ref{eq:ub2}) can be exploited in the double-limit model to
derive a tighter (asymptotic) upper-bound. As expected, the answer
is negative: indeed, denoting the time limit $f(n)=\limsup_T
f(n,T)$, a tighter bound could only be derived if
\begin{equation*}
1-f(n)=o(1)~.
\end{equation*}
This holds, however, only when $\lambda(n)=0$, which is a
degenerate case.

\subsubsection{Lower bounds}
By explicitly constructing a routing and scheduling scheme, the
underlying idea to compute lower bounds on $\lambda(n)$ is based
on the condition
\begin{equation}
\lambda(n) l\leq c~,\label{eq:lb}
\end{equation}
where $l$ is an \textit{upper bound} on the number of e2e
transmissions a node \textit{needs} to act as a relay, whereas $c$
is the maximal rate at which a node \textit{can} transmit (see
p.~400, $2^{\textrm{nd}}$ column, $1^{\textrm{st}}$ equation
from~\cite{Gupta00}). For the network model from~\cite{Gupta00},
$l=\Theta\left(\sqrt{n\log{n}}\right)$ and $c=\Theta(1)$.

Alike the upper-bound, the lower-bound continues to hold in a
finite-space model under appropriate structural properties. In a
finite-space/time model, however, the lower bound ceases to hold.
For a counterexample (relative to current conditions), consider
$3$ nodes numbered as $\{1,2,3\}$, the (direct) source-destination
pairs $\left\{(1-2), (2-3), (3-1)\right\}$, and assume that all
transmissions interfere with each other. Fitting Eq.~(\ref{eq:lb})
yields $l=1$, $c=\frac{1}{3}$, and thus $\lambda(3)=\frac{1}{3}$.
Evidently, this rate can only be sustained for specific values of
$T$ (e.g., if exogenous arrivals occur at times $3s+1$ at all
nodes, and under a round-robin scheduling, then the lower bound
only holds at times $T=3s$ for $s\geq0$).

In finite scenarios in which the lower-bound does hold, it can be
further tightened using the same multiplicative term as for the
upper bound, i.e.,
\begin{equation} \lambda(n)l\left(1-f(n,T)\right)\leq
c~.\label{eq:lb2}
\end{equation}
Note that the effective multiplicative factor for the lower bound
is in fact $\frac{1}{1-f(n,T)}$. A related observation is that, as
expected, (asymptotic) tighter bounds can only be obtained when
$1-f(n)=o(1)$, i.e., a degenerate case.

In conclusion, the upper and lower bounds arguments from
Eqs.~(\ref{eq:gkar}) and~(\ref{eq:lb}) hold immediately in a
finite-space model. In finite-space/time models, only the former
holds in general, and both can be improved by a multiplicative and
fractional, respectively, factor $1-f(n,T)$; next we will show
that this factor can be quite small (and thus detrimental),
including at large values of $n$. Finally, we remark that even by
considering a tightening factor, as in Eqs.~(\ref{eq:ub2}) and
(\ref{eq:lb2}), capacity results remain restricted to first
moments (time averages) only. These limitations demand thus for
`richer' capacity results in terms of probability distributions,
which can readily render \textit{all} moments, and more generally
for alternative analytical techniques beyond the convenient but
simplistic averages-based calculus.

\subsection{Simulations for $1-f(n,T)$}\label{sec:simus}
Here we simulate the multiplicative/fractional factor $1-f(n,T)$
identified in Eqs.~(\ref{eq:ub2}) and (\ref{eq:lb2}). For the
clarity of the exposition, we consider both a simple setting,
consisting of a single e2e transmission along a line network, and
a more involved random network.

\subsubsection{Example 1: A Single E2E Transmission}\label{sec:e1}
\begin{figure}[h]
\vspace{0cm}
\begin{center}
\shortstack{
\includegraphics[width=0.42\linewidth,keepaspectratio]{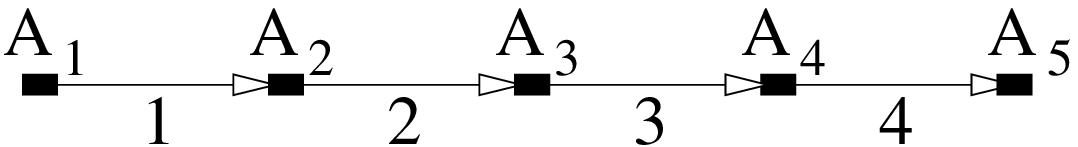}
\\
{\footnotesize (a) A line network} } \shortstack{\hspace{2cm}
\includegraphics[width=0.22\linewidth,keepaspectratio]{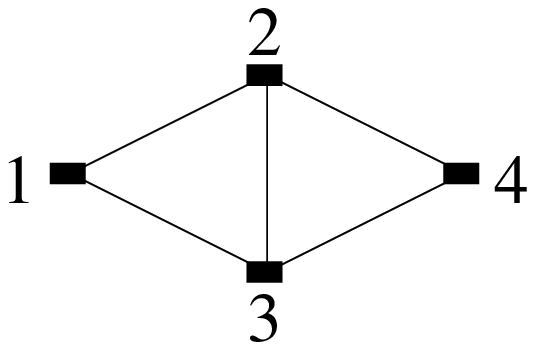}
\\
{\footnotesize (b) Contention graph} } \caption{A multi-hop
network and its contention graph} \label{fig:5Nodes}
\end{center}
\end{figure}

Consider a multi-hop network with $n$ nodes, of which a source
node $A_1$ transmits (packets) to a destination node $A_n$ using
the relay nodes $A_i$, $i=2,3,\dots,n-1$. The (single) end-to-end
(e2e) transmission is denoted by $[A_1\rightarrow A_n]$. We denote
by $c_r$ the contention range of link $i$, i.e., link $i$
interferes with any link $j$ satisfying $|i-j|<c_r$. An example
for $n=5$ is shown in Figure~\ref{fig:5Nodes}.(a). A corresponding
contention graph for $c_r=3$ is shown in
Figure~\ref{fig:5Nodes}.(b). Here, each vertex $i$ stands for the
uni-directional transmission $[A_i\rightarrow A_{i+1}]$,
$i=1,2,3,4$, and there is a link between nodes $i$ and $j$ if the
corresponding transmissions interfere with each other; according
to this contention graph, links $1$ and $4$ can simultaneously and
successfully transmit.

Next we illustrate the average percentage of non-empty buffers
$1-f(n,T)$ for the network setting from Figure~\ref{fig:5Nodes}
over a time span $T$, and for two MAC protocols: (slotted) Aloha
and CSMA/CA, to be described in Section~\ref{sec:mac}.

\begin{figure}[h]
\vspace{0cm}
\begin{center}
\shortstack{\hspace{0cm}
\includegraphics[width=0.4\linewidth,keepaspectratio]{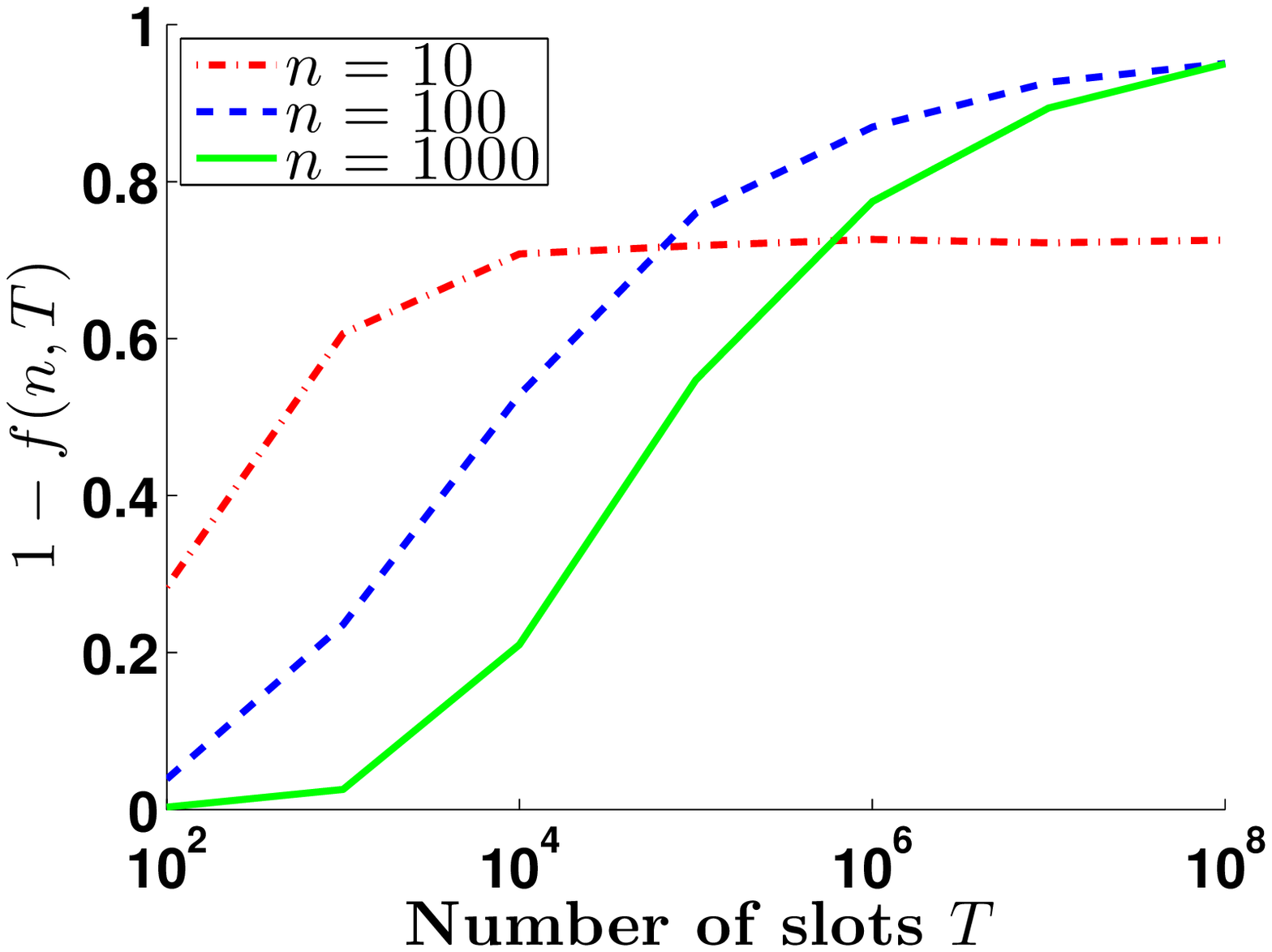}
\\
{\footnotesize (a) Aloha} } \shortstack{\hspace{2.0cm}
\includegraphics[width=0.4\linewidth,keepaspectratio]{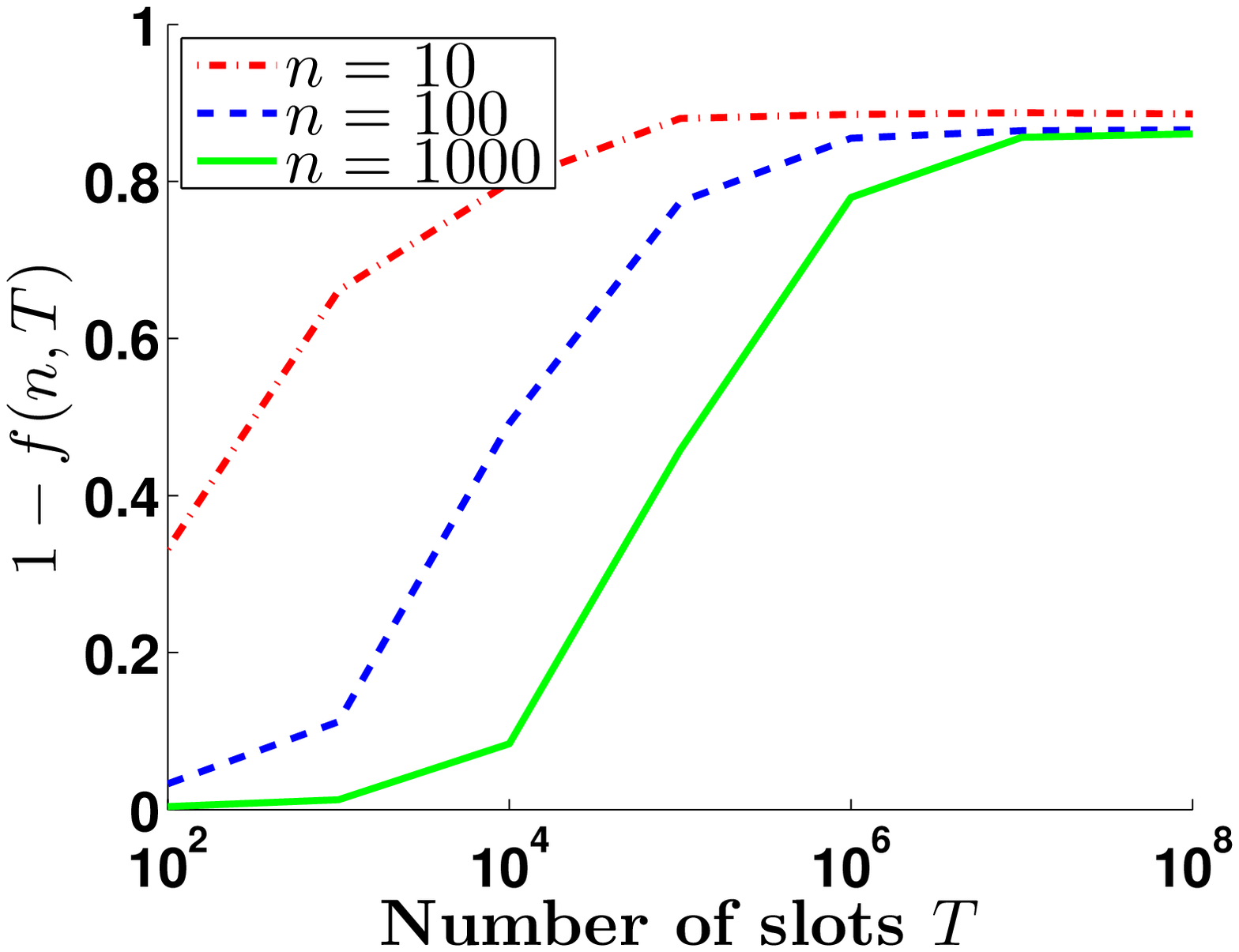}
\\
{\footnotesize (b) CSMA/CA} } \caption{The average percentage of
non-empty buffers $1-f(n,T)$ as a function of time $T$ in a line
network} \label{fig:line}
\end{center}
\end{figure}

Figure~\ref{fig:line}.(a) illustrates the Aloha case. The network
capacity is $\lambda(n)=p(1-p)^4$, where $p$ is the transmission
probability; by optimizing, $p=0.2$ and $\lambda(n)\approx 0.08$,
which is the rate injected at the first node (recall that there is
a single e2e transmission). We observe that for small number of
nodes (e.g., $n=10$), the (average) percentage of empty buffers
remain significant, even when taking the limit in $T$. This effect
is due to insufficient amount of spatial reuse. For larger number
of nodes, however, there is sufficient amount of spatial reuse and
the percentage of empty buffers goes to zero. This convergence,
however, is surprisingly slow (at $T=10^8$ there are still
$\approx 5\%$ empty buffers).

Figure~\ref{fig:line}.(b) illustrates the CSMA/CA case with
average backoff and transmission times $\nu^{-1}=10$ and
$\mu^{-1}=10$, respectively. Because a formula for $\lambda(n)$ is
difficult to obtain, unlike in the Aloha case, we numerically
searched for the minimum value of $\lambda(n)$ such that the total
amount of packets in all buffers, except destinations, at any
time, is smaller than $10^6$ over a maximum time span $T=10^8$. We
obtained $\lambda(n)\approx 0.17$ for $n=10,100,1000$. As for
Aloha, the homogeneity is due to the dominant effect of a
bottleneck; unlike Aloha, however, CSMA/CA is subject to
transmission correlations spanning the entire network. Relative to
the Aloha case, we also remark a sharper rate of decrease of
$f(n,T)$; however, this rate slows down earlier (e.g., for $n=100$
there are still $\approx 13\%$ empty buffers in contrast to only
$\approx 5\%$ for Aloha).

\subsubsection{Example 2: A Random Network}\label{sec:e1}
We now consider a closer network setting to the one
from~\cite{Gupta00}. We first randomly place $n$ nodes on a square
and then randomly choose a destination for all sources
$1,2,\dots,n$; for both random generations we use uniform
distributions. Then we determine the minimum transmission range
such that e2e paths exist for all source-destination pairs; these
paths are constructed using a shortest path algorithm, where the
weight of each link is set to one. Each node stores the locally
generated and incoming packets in a FIFO buffer.

As in Example 1, we illustrate $1-f(n,T)$ as a function of $T$ for
both Aloha and CSMA/CA. For Aloha we set the nodes' transmission
probability as the inverse of the maximum node-degree amongst all
nodes. Moreover, we use the same numerical search form Example 1
to determine the capacity $\lambda(n)$ for both Aloha
($\lambda(10)\approx 0.01$, $\lambda(100)\approx 0.002$, and
$\lambda(1000)\approx 0.0004$) and CSMA/CA ($\lambda(10)\approx
0.03$, $\lambda(100)\approx 0.007$, and $\lambda(1000)\approx
0.0009$).

\begin{figure}[h]
\vspace{-.15cm}
\shortstack{\hspace{0cm}
\includegraphics[width=0.45\linewidth,keepaspectratio]{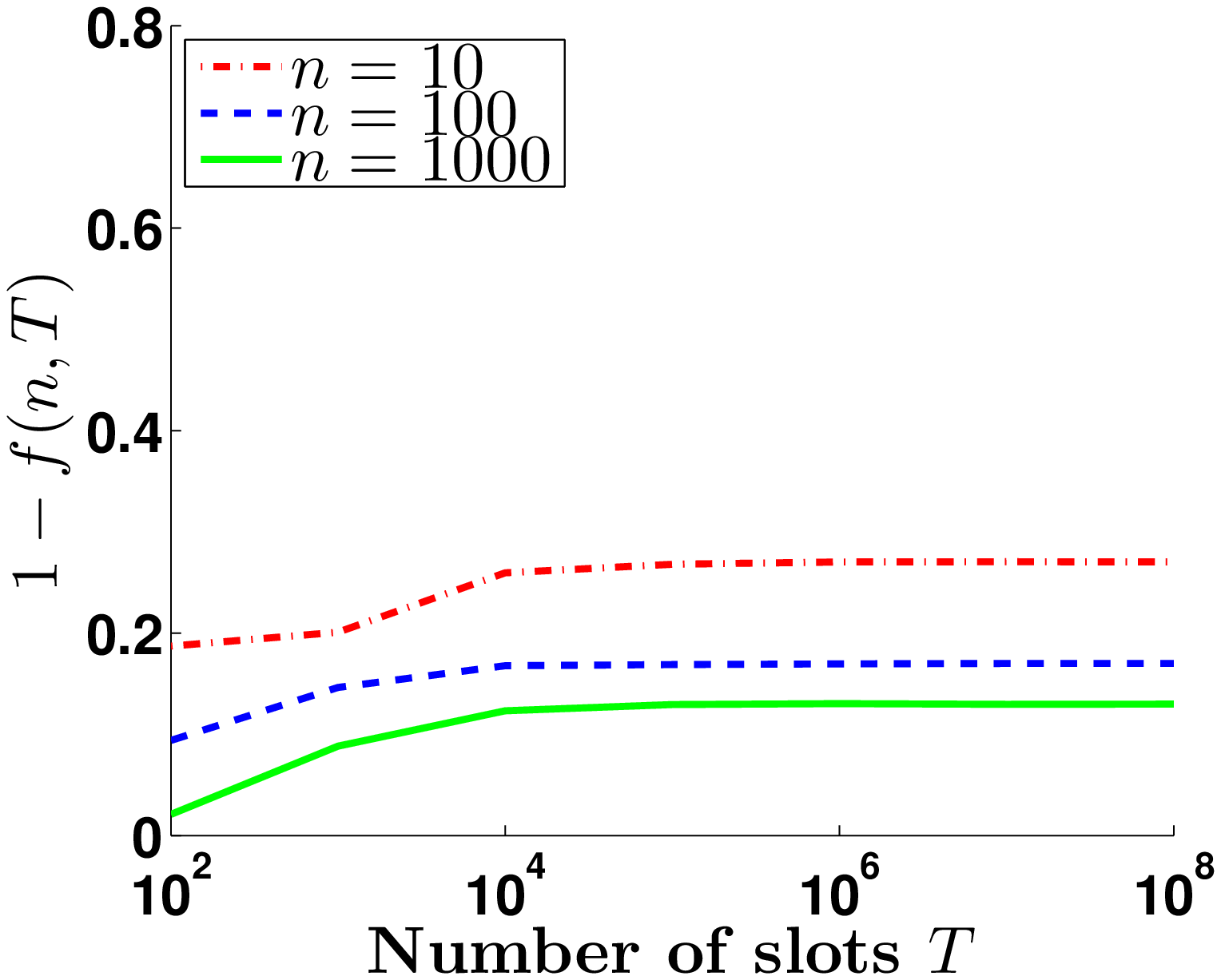}
\\
{\footnotesize (a) Aloha} } \shortstack{\hspace{1cm}
\includegraphics[width=0.45\linewidth,keepaspectratio]{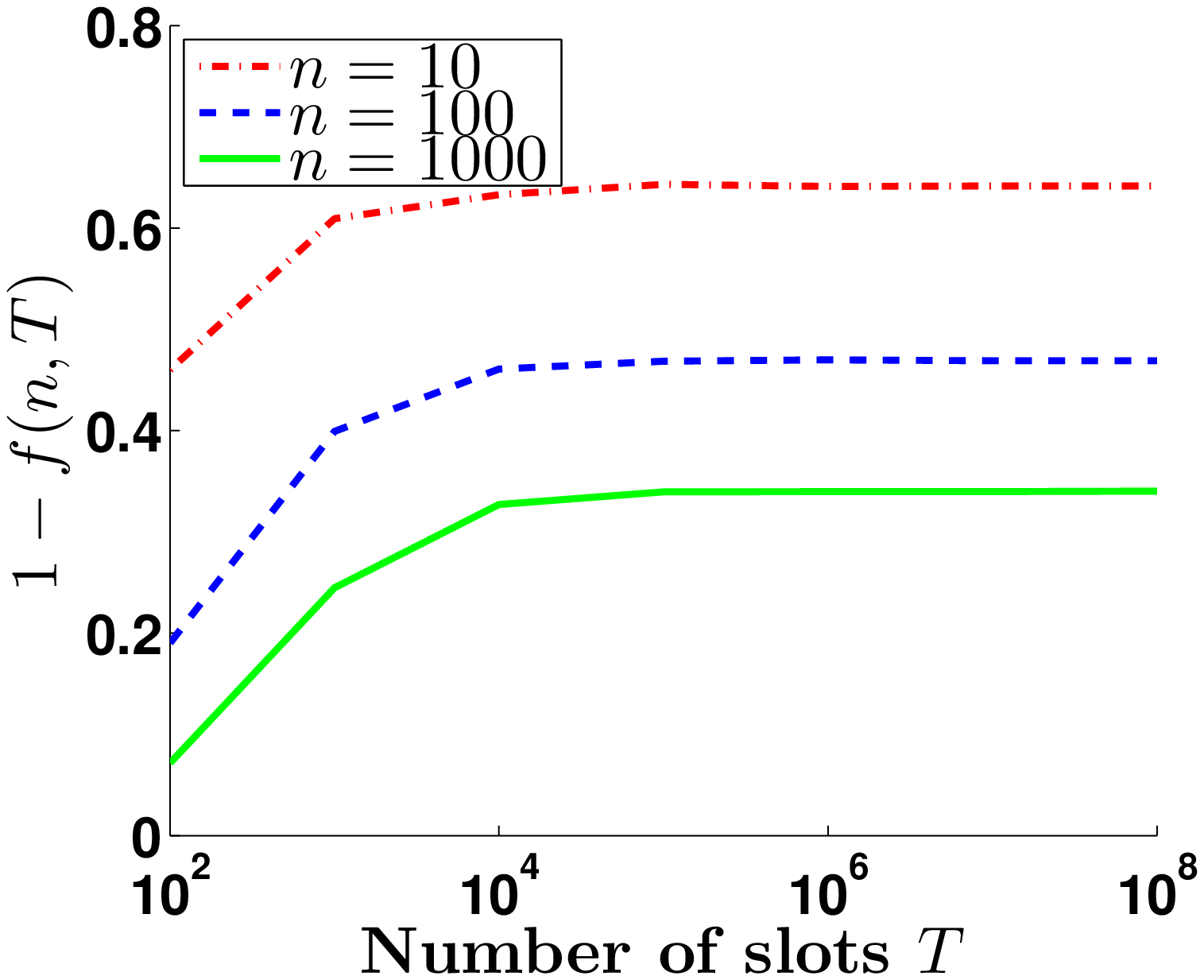}
\\
{\footnotesize (b) CSMA/CA} } \caption{The average percentage of
non-empty buffers $1-f(n,T)$ as a function of time $T$ in a random
network} \label{fig:random}
\end{figure}

From Figure~\ref{fig:random}.(a) we observe a clear convergence of
$1-f(n,T)$; the perhaps surprisingly low limits are conceivably
due to the homogeneous transmission probabilities accounting for
the bottleneck region. Unlike in the line network, there is a
consistent monotonic behavior in the number of nodes $n$, which is
likely due to the more uniform structure of the random network
setting. In the CSMA/CA case, Figure~\ref{fig:random}.(b)
illustrates that much fewer buffers (by a factor of roughly three)
are empty than in the Aloha case, which suggest a less burstier
behavior in CSMA/CA.

Clearly, Figures~\ref{fig:line} and \ref{fig:random} open several
fundamental questions on network queueing behavior for Aloha and
CSMA/CA, which may help improving the two. Their main purpose,
however, is to convincingly show that the average percentage of
non-empty buffers $1-f(n,T)$ is quite small especially in random
networks, and in general at small time scales. The key observation
is the monotonic (decreasing) behavior (excepting the special
Aloha line with $n=10$) in the number of nodes $n$. This behavior
is `somewhat expected' in large networks, by invoking laws of
large numbers arguments, i.e., the overall incoming and outgoing
flows tend to stabilize and thus buffers tend to decrease. This
means that both the original upper and lower bounds from
Eqs.~(\ref{eq:gkar}) and~(\ref{eq:lb}) become conservative in
asymptotic regimes (in $n$). Evidently, this observation is
relative to the settings herein herein; whether it generally holds
requires further analysis.

In conclusion, the results from this section motivate the need for
an analytical approach for network capacity in finite regimes. At
this point, we ought to be rigorous in defining capacity in finite
time. Concretely, given a time $t$ and an arrival process $D(t)$
at the destination of an e2e path, we are interested in bounds of
the form
\begin{equation*}
\P\left(D(t)\leq\lambda_t t\right)\leq\eps~,
\end{equation*}
for some violation probability $\eps$. Here, $\lambda_t$ is a
lower bound on the throughput (capacity) rate of the e2e
transmission; upper bounds can be defined similarly by changing
the inequality in the probability event.

\section{A System-Theoretic Approach for Finite Time Capacity}\label{sec:worksol}
According to the previous discussion, the main challenge to derive
capacity results in terms of distributions, and also in finite
regimes, is accounting for queueing behavior in a conceivably
non-Jackson queueing network. In particular, it is especially hard
to analytically keep track of buffer occupancies at relay nodes.

To address this problem, we next describe a general solution to
circumvent the characterization of buffer occupancies at the relay
nodes, by making an analogy with LTI systems. The idea is to view
a single-hop transmission as follows: the data at the source and
destination stand for the input and output signals, respectively,
whereas the transmission and its characteristics, accounting for
both data unavailability due to burstiness or noise due to
interference, are modelled by `\textit{the system}' transforming
the input signal; as expected, this \textit{system} is not linear,
but there is a subtle analogy with LTI systems which drives its
analytical tractability.

To present the main idea in an approachable manner, from the point
of view of notational complexity, we focus on the simplified
simple line network from Figure~\ref{fig:5Nodes}; extensions to
more involved topologies will be mentioned as well.

\begin{figure}[h]
\centerline{\epsfig{figure=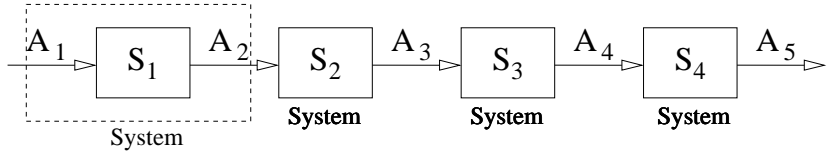,width=4.3in,keepaspectratio}
} \caption{A system interpretation of the multi-hop network from
Figure~\ref{fig:5Nodes}} \label{fig:5System}
\end{figure}

Figure~\ref{fig:5System} illustrates a system view for the e2e
transmission from Figure~\ref{fig:5Nodes}. With abuse of notation
the $A_i$'s stand for the input/output signals, and the $S_i$'s
stand for the impulse-responses of the systems. The key property
of the impulse-responses is to relate the input and output signals
through a convolution operation, i.e.,
\begin{equation}
A_{i+1}=A_i\conv S_i~.\label{eq:convolution}
\end{equation}
As it will become more clear in Section~\ref{sec:mac}, the
convolution operation denoted here by the symbol `$\conv$'
operates in a $(\textrm{min},+)$ algebra. To be more specific, let
$A_i$ stand for a stochastic process $A_i(t)$, which counts the
number of packets in the time interval $[0,t]$ at node $i$; also,
let $S_i$ stand for (some) bivariate stochastic processes
$S_i(s,t)$. Then the $(\min,+)$ convolution operation expands as
$$A_{i+1}(t)=\min_{0\leq s\leq
t}\left\{A_i(s)+S_i(s,t)\right\}~\forall t\geq0~.$$

\begin{figure}[h]
\centerline{\epsfig{figure=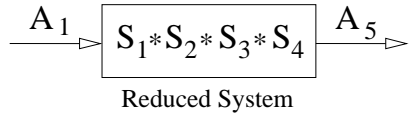,width=2.1in,keepaspectratio}
} \caption{The composition of the four systems from
Figure~\ref{fig:5System} into a single system} \label{fig:1System}
\end{figure}

The relationship from Eq.~(\ref{eq:convolution}) has two key
properties. One is that it holds for \textit{any} input signal
$A_i$, which is hard to derive at relay nodes (when $i\geq2$). In
other words, the impulse-response $S_i$ \textit{entirely}
characterizes the \textit{system}, i.e., the single-hop
transmission $i$, which is a key feature of LTI systems since it
enables their analytical tractability. The convolution operation
has also the useful algebraic property of \textit{associativity}.
The two properties circumvent keeping track of $A_i(t)$ at the
relay nodes (i.e., for $i=2,3,4$). Indeed, by applying
associativity, and using the physical property that the output
signal in a system is the input signal at the downstream system,
the composition of the four systems from Figure~\ref{fig:5System}
yields the reduced system from Figure~\ref{fig:1System}.

The reduced system dispenses with the intermediary signals $A_2$,
$A_3$, and $A_4$, and instead it retains the impulse-responses in
a composition (or e2e) form, i.e.,
\begin{equation}
A_5=A_1\conv\left(S_1\conv S_2\conv S_3\conv
S_4\right)~.\label{eq:e2econv}
\end{equation}

What has yet to be shown concerns the existence of (analytical)
expressions for the impulse-responses $S_i$'s, satisfying the key
property from Eq.~(\ref{eq:convolution}). The other open issue is
whether the convenient reduction from Figure~\ref{fig:1System} and
Eq.~(\ref{eq:e2econv}) is analytically tractable. We will next
show that impulse-responses can be constructed for single-hop
links in an analogous manner as in LTI systems, depending on the
underlying MAC protocol, whereas analytical tractability follows
the steps of large deviations or stochastic network calculus
theories.


\section{Markov Modulated Transmission Processes
(MMTPs)}\label{sec:mac} Wireless networks must deal with the
fundamental interference problem: two simultaneous transmissions
may jointly fail if they interfere with each other. MAC protocols
partially resolve this problem by reducing the number of
collisions and consequently increasing the network capacity.
Obviously, different MAC protocols can lead to different
capacities.

To capture the influence of MAC protocols on the throughput
capacity, we introduce the concept of Markov Modulated
Transmission Process (MMTP). An MMTP is defined for each link, and
models the link's activities (successful/unsuccessful
transmissions and idle periods) as a time process and according to
the workings of the underlying MAC protocol. The model consists of
a Markov chain/process $X(t)$ (depending on the underlying
discrete/continuous time model), where $t$ is a time parameter,
which \textit{modulates} the transmission rate of a link
$[i\rightarrow j]$, if the source $i$ has data to send at time
$t$. In discrete time, the transmission rate in a slot $t$ is
\begin{equation}
S(t-1,t)=C_{X(t)}~,\label{eq:smmtp}
\end{equation}
where $C_{X(t)}$ is the \textit{Markov Modulated Transmission
Process} defined on the state space ${\mathcal{T}}$ of the Markov
chain $X(t)$. It is described as
\begin{equation}
C_{X(t)}=\left\{\begin{array}{lll}C&,&X(t)\in{\mathcal{T}}_{[i\rightarrow
j]}\\0&,&\textrm{otherwise~,}\end{array}\right.\label{eq:mmtp}
\end{equation}
where ${\mathcal{T}}_{[i\rightarrow j]}\subseteq{\mathcal{T}}$
denotes the set of \textit{favorable} states of $X(t)$ for the
link $[i\rightarrow j]$, which would guarantee a successful
transmission if the link has data to send at time $t$; whenever a
transmission is successful we assume a constant throughput
capacity $C$. The rest of the states
$X(t)\in{\mathcal{T}}\backslash{\mathcal{T}}_{[i\rightarrow j]}$
model the times when the link attempts an unsuccessful
transmission or it is idle in accordance to the MAC protocol.

The MMTP process $C_{X(t)}$ defined in Eq.~(\ref{eq:mmtp}) is
modulated by the Markov chain $X(t)$, and it is conceptually
identical with, e.g., Markov Modulated Poisson Processes (MMPPs),
which have been used in teletraffic theory to model voice or video
(Heffes and Lucantoni~\cite{Heff8609:Markov}). $X(t)$ can be
defined either for the \textit{whole} network (when it modulates
the transmission opportunities of \textit{all} the links) or for
\textit{each} link separately. In turn, $C_{X(t)}$ is always
separately defined for \textit{each} link.

We point out that the process $S(s,t)$, which we loosely
introduced in Eq.~(\ref{eq:smmtp}) through its increments
$S(t-1,t)$, directly corresponds to the \textit{impulse-response}
process introduced in Section~\ref{sec:worksol} to
\textit{entirely characterize} the behavior of a single-hop
transmission in system theoretic terms (see
Figure~\ref{fig:5System} and Eq.~(\ref{eq:convolution})). We
mention that the impulse-response defined in Eq.~(\ref{eq:smmtp})
corresponds to the \textit{effective capacity} concept proposed by
Wu and Negi in~\cite{WuNegi03} to model the instantaneous channel
capacity. This concept was used by Tang and Zhang~\cite{Tang07} to
analyze the impact of physical layer characteristics (e.g., MIMO)
on delay at the data-link layer. The MMTP idea was also used
explicitly by Fidler~\cite{FidlerFading06},
Mahmood~\et~\cite{Mahmood11}, Al-Zubaidy~\et~\cite{Zubaidy13},
Zheng~\et~\cite{Zheng13} and implicitly by
Ciucu~\et~\cite{CiHoHu10,CiucuISIT11}, to model channel service
processes with Markov chains. Relative to these previous works,
our contribution herein is to fit the effective capacity concept,
defined as a Markov modulated process, for three specific MAC
protocols in the framework of the stochastic network calculus. A
key feature of network calculus is that it facilitates the
analysis of network queueing problems by relying on a subtle
analogy with LTI systems~(see Le Boudec and
Thiran~\cite{Book-LeBoudec}, and Ciucu and
Schmitt~\cite{Ciucu12}).

In the following we explicitly construct the impulse-response
process $S(s,t)$, and the underlying Markov process $X(t)$ and
MMTP $C_{X(t)}$, for three MAC protocols. In addition, we outline
the key steps to compute the (per-flow) distribution of the
throughput capacity in closed-form.

\subsection{Centralized scheduling} Assuming a time-slotted model
and the nodes' perfect synchronization, the idea of centralized
scheduling is to pre-allocate the transmission slots to the nodes
in order to avoid collisions. In unsaturated scenarios, an optimal
solution (i.e., attaining maximal throughput) would require
significant overhead as the centralized scheduler would require
keeping track of the arrival processes at each of the nodes. Even
in saturation scenarios, the optimality problem is in fact
NP-complete in general networks (see, e.g.,
Sharma~\et~\cite{SharmaInf07}).

For the network model from Figure~\ref{fig:5Nodes}.(a-b), the
optimal scheduling allocation starting from slot $1$, in terms of
links, is: $\{1,2,3,(1,4),2,3,(1,4),\dots\}$; for instance, link
$2$ is allocated the slots $2,5,8,\dots$ That means that link $2$
is given full transmission capacity (say $C$) during these slots,
which suggests that the bivariate function
\begin{equation}
S_2(s,t)=C\sum_{u=s+1}^tI_{(u-2)\%3=0}~\forall 2\leq s\leq
t~,\label{eq:link2}
\end{equation}
and $0$ everywhere else, characterizes the capacity of link $2$ in
terms of the $(\textrm{min},+)$ convolution from
Eq.~(\ref{eq:convolution}) ($I_{E}$ denotes the indicator function
taking the values $0$ and $1$, depending on whether the event $E$
is false or true). The intuition is that $S_2(s,t)$ counts the
number of packets transmitted over link $2$ in the time interval
$(s,t]$, if node $2$ is saturated.

This saturation condition translates into system theoretic terms
as follows: the input signal to the second system in
Figure~\ref{fig:5System} is the infinite signal $A_2(t)=\infty$
for $\forall t$ (also called the \textit{impulse}), whereas the
corresponding output, i.e., the \textit{impulse-response}, is the
signal $S_2(t)$, or $S_2(0,t)$ in the notation from
Eq.~(\ref{eq:link2}). Therefore, the construction of $S_2(s,t)$ is
analogous to the construction of impulse-response functions in LTI
systems, which are the output from an LTI system with input given
by the Kronecker signal (see~\cite{LeeVaraiya03}). Although the
system representing the link's transmission is \textit{not}
linear, even under the $(\min,+)$ algebra, the constructed process
$S_2(s,t)$ \textit{entirely characterizes} link $2$, i.e.,
\begin{equation}
A_3(t)=\min_{0\leq s\leq t}\left\{A_2(s)+S_2(s,t)\right\}~\forall
t\geq0~,\label{eq:conv23}
\end{equation}
for \textit{all} $A_2(t)$ at the input of the second system.

\begin{figure}[t]
\vspace{-.15cm}
\begin{center}
\shortstack{\hspace{-0.0cm}
\includegraphics[width=0.25\linewidth,keepaspectratio]{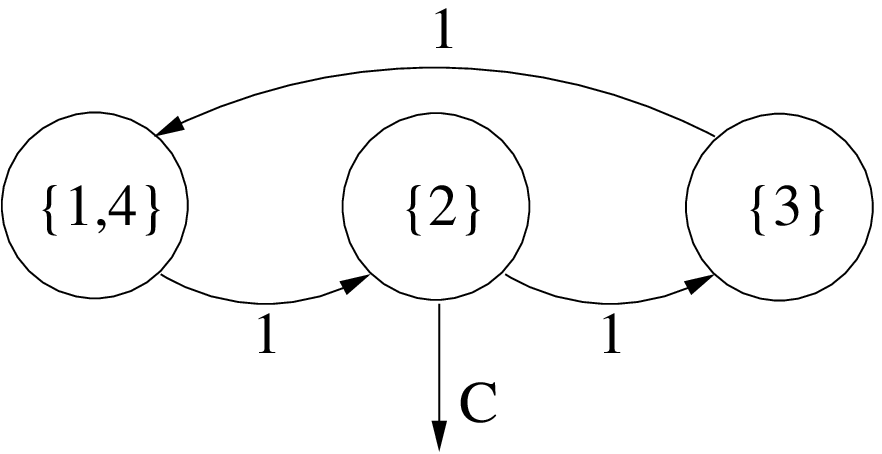}
\\
{\footnotesize (a) Centralized scheduling} }
\shortstack{\hspace{4cm}
\includegraphics[width=0.16\linewidth,keepaspectratio]{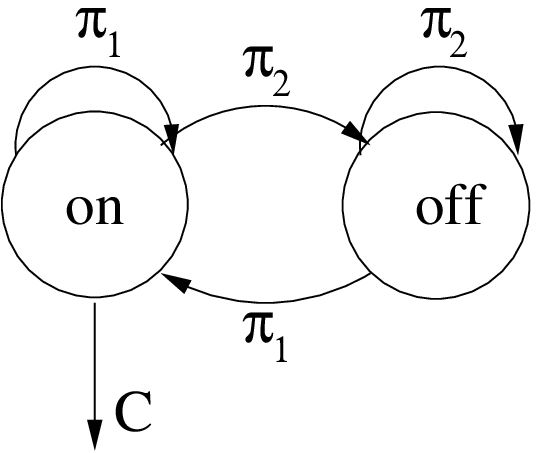}
\\
{\footnotesize~~~~~~~~~(b) Aloha} } \caption{Markov Modulated
Transmission Processes (MMTPs) for link $[A_2\rightarrow A_3]$}
\label{fig:centralAloha}
\end{center}
\end{figure}

So far we directly constructed $S_2(s,t)$ without resorting on an
MMTP $C_{X(t)}$. The underlying MMTP, and also the modulating
Markov chain $X(t)$, are depicted in
Figure~\ref{fig:centralAloha}.(a). The states of $X(t)$ denote the
set of transmitting links (according to the centralized schedule).
The transition probabilities between the states are all equal to
$1$, thus reflecting the deterministic nature of centralized
scheduling. The MMTP process for link $[A_2\rightarrow A_3]$ is
\begin{equation*}
C_{X(t)}=\left\{\begin{array}{lll}C&,&\textrm{if}~X(t)=\left\{2\right\}\\
0&,&\textrm{otherwise~.}\end{array}\right.
\end{equation*}
The MMTPs for the other links are defined similarly; for instance,
for links $1$ and $4$, the only change is that $C_{X(t)}=C$ when
$X(t)=\left\{1,4\right\}$. Note that all MMTPs share the same
Markov chain modulating the transmission opportunities at the
network level. Moreover, $X(t)$ and $C_{X(t)}$ jointly reproduce
the expressions of the impulse responses (e.g., of $S_2(s,t)$ from
Eq.~(\ref{eq:link2})) according to the definition from
Eq.~(\ref{eq:smmtp}).

Concerning analytical tractability, we remark that the reduced
system from Figure~\ref{fig:1System} is implicitly tractable since
the constructed impulse-responses $S_i(s,t)$ are deterministic
functions. The overall impulse-response of the e2e path
$S=S_1\conv S_2\conv S_3\conv S_4$ from Eq.~(\ref{eq:e2econv}) can
be computed directly, i.e.,
\begin{equation}
S(s,t)=C\sum_{u=s+1}^tI_{(u-1)\%3=0}~\forall 1\leq s\leq
t~,\label{eq:cse2e}
\end{equation}
and $0$ everywhere else. To quickly check how the e2e
impulse-response $S$ captures the causality condition, assume that
$A_1(1)=1$ (i.e., one packet arrives at time $1$). Then this
packet departs the network no sooner than at time $4$: indeed,
$A_5(3)=A_1\conv S(3)=0$, as $S(0,3)=0$, and that
$A_5(4)=\min_{0\leq s\leq 4}\left\{A_1(s)+S(s,4)\right\}=1$, with
the minimum being attained at $s=0$.

Although the constructions of the MMTP's above is not technically
necessary, as the impulse-responses were directly constructed, and
the impulse-response $S(s,t)$ of the e2e path could be in
principle determined by other means than computing an e2e
convolution, we regard this detour to be insightful for the
construction of impulse-responses for the more challenging cases
of Aloha and CSMA/CA protocols.


\subsection{Aloha}\label{sec:aloha} The slotted-Aloha MAC protocol (called
Aloha here) is an elegant solution to circumvent centralized
scheduling (see Ambramson~\cite{Abramson70}). The key idea is that
each node attempts to transmit with some probability in each time
slot and when data is available; a transmission $[i\rightarrow j]$
is successful in some time slot $t$ if $i$ is the only node in the
interference range of node $j$ attempting to transmit in that
slot. While the protocol is entirely distributed, it may
experience significant performance decay, e.g., the achieved
capacity can be as small as $36\%$ of the theoretical limit.

To construct the impulse-response processes for the line network
from Figure~\ref{fig:5Nodes}.(a-b), we first construct the MMTP
processes. We focus again on link $[A_2\rightarrow A_3]$. The
underlying MMTP, and also the modulating Markov chain $X(t)$, are
depicted in Figure~\ref{fig:centralAloha}.(b). The meaning of
state `on' is that, while $X(t)$ delves in it, the relay node
$A_2$ successfully transmits (if there is data to send). In turn,
while $X(t)$ delves in state `off', $A_2$ is either idle (in
accordance with the workings of Aloha) or it is involved in a
collision. Assuming, for convenience, that all nodes transmit with
the same probability $p$, the transition probabilities are
$\pi_1=p(1-p)^3$ and $\pi_2=1-\pi_1$ (the power of $3$ is the
degree of node $2$ in the contention graph from
Figure~\ref{fig:5Nodes}.(b)). For this Markov chain, the
steady-state probabilities are $\pi_{\textrm{on}}=\pi_1$ and
$\pi_{\textrm{off}}=\pi_2$, and $X(t)$ has the convenient property
of statistically independent increments, e.g., $\forall t$
\begin{equation}
\P\left(X(t+1)=\textrm{`on'}|X(t)=\textrm{`off'}\right)=\P\left(X(t+1)=\textrm{`on'}\right)~.\label{eq:indincr}
\end{equation}

The definition of the associated MMTP should be intuitive at this
point, i.e.,
\begin{equation}
C_{X(t)}=\left\{\begin{array}{lll}C&,&\textrm{if}~X(t)=\textrm{`on'}\\
0&,&\textrm{otherwise~,}\end{array}\right.\label{eq:mmtpAloha}
\end{equation}
as also illustrated in Figure~\ref{fig:centralAloha}.(b). In other
words, $A_2$ can successfully transmit (assuming it has data) at
full rate $C$ while the modulating process $X(t)$ delves in the
favorable state `on'.

The construction of the other links' MMTP processes is almost
identical, except for the transmission probabilities of the
modulating process. For instance, for the links $[A_1\rightarrow
A_2]$ and $[A_4\rightarrow A_5]$, the new transmission
probabilities are $\pi_1=p(1-p)^2$ and $\pi_2=1-\pi_1$ (the power
of $2$ is the common degree of nodes $1$ and $4$ in the contention
graph from Figure~\ref{fig:5Nodes}.(b)).

These MMTP processes directly determine the impulse-response
functions $S_i(s,t)$, corresponding to the single-hop links
$i=1,2,3,4$, according to the definition from
Eq.~(\ref{eq:smmtp}). Furthermore, the composition property of the
$S_i(s,t)$'s in the underlying $(\min,+)$ algebra lends itself to
the \textit{entire characterization} of the throughput capacity
over the e2e path as in Eq.~(\ref{eq:e2econv}), i.e.,
\begin{equation*}
A_5=A_1\conv S~,\label{eq:e2econvr}
\end{equation*}
where $S=S_1\conv S_2\conv S_3\conv S_4$ is the impulse-response
of the e2e path.

Unlike centralized scheduling which may lend itself to an explicit
expression for $S$ (e.g., as in Eq.~(\ref{eq:cse2e})), Aloha is
conceivably more challenging with respect to the analytical
tractability of the reduced system. One immediate issue lies in
the probabilistic structure of the local impulse-responses
$S_i$'s. A more subtle issue lies in the fact that the $S_i$'s are
statistically correlated random processes, even in the simplified
line network.

To deal with these challenges, the key idea is to \textit{trade
analytical exactness for tractability}. More concretely, instead
of \textit{exactly} deriving the e2e transient capacity in
closed-form (an open problem in itself), we compute bounds by
relying on large deviation techniques (e.g., as
in~\cite{CiHoHu10}). Let us illustrate such computations for the
first two hops only, and a saturation assumption at node $A_1$
(the relay nodes are however \textit{not} assumed to be
saturated). The probability of violating a lower bound
$\lambda_t$, on the \textit{transient throughput rate} over the
time scale $[0,t]$, can be computed as follows for some $\theta>0$
\begin{eqnarray}
&&\hspace{-1.3cm}\P\left(A_3(t)\leq \lambda_t t\right)=\P\left(S_1\conv S_2(t)\leq \lambda_t t\right)\label{eq:derivationAloha}\\
&&\hspace{-0.50cm}=\P\left(\sup_{0\leq s\leq t}\left\{\lambda_t
t-S_1(s)-S_2(s,t)\right\}\geq 0\right)\notag\\
&&\hspace{-0.50cm}\leq\sum_{0\leq s\leq t}e^{\theta\lambda_t
t}E\left[e^{-\theta S_1(s)}\right]E\left[e^{-\theta
S_2(s,t)}\right]~,\notag
\end{eqnarray}
by using Boole's inequality\footnote{For some probability events
$E$ and $F$, Boole's inequality states that $\P\left(E\cup
F\right)\leq\P(E)+\P(F)$.} and the Chernoff bound\footnote{For
some r.v. $X$, $x\in\R$, and $\theta>0$, the Chernoff bound states
that $\P(X>x)\leq E\left[e^{\theta X}\right]e^{-\theta x}$.}; in
the first line we also used the saturation assumption at mode
$A_1$, e.g., $A_1(1)=\infty$. The last step is also based on the
statistical independence between the impulse-responses $S_1(u,s)$
and $S_2(s,t)$ (in order to apply $E[XY]=E[X]E[Y]$ for some
independent r.v.'s $X$ and $Y$); this holds because $(u,s]$ and
$(s,t]$ are non-overlapping intervals, whereas the corresponding
Markov modulated processes of $S_1(u,s)$ and $S_2(s,t)$ have
statistically independent increments (recall the
\textit{convenient} property from Eq.~(\ref{eq:indincr})).
Therefore, although $S_1(u,s)$ and $S_2(u,s)$ are correlated over
overlapping intervals, the expansion of the $(\min,+)$ convolution
(in terms of non-overlapping intervals) and the independent
increments property from Eq.~(\ref{eq:indincr}) justify the last
step.

The Laplace transforms in the last equation can be computed
explicitly for the impulse-responses. Concretely, one has
$E\left[e^{-\theta S_i(s,t)}\right]\leq e^{-\theta r_s(t-s)}$,
where $r_s=\frac{\log{\left(qe^{-\theta
C}+1-q\right)^{-1}}}{\theta}$ and $q=p(1-p)^2$. Evaluating the
nested sums from Eq.~(\ref{eq:derivationAloha}) yields the
following closed-form bound
\begin{equation}
\P\left(A_3(t)\leq \lambda_t
t\right)\leq\inf_{\theta}(t+1)e^{-\theta(r_s-\lambda_t)t}~.\label{eq:lbound}
\end{equation}
The result is not explicit, as $\theta$ is to be optimized
(typically numerically). Setting the right-hand side to some
violation probability, the lower bound $\lambda_t$ on the
transient capacity rate follows immediately. We mention that
probabilistic upper bounds can also be derived similarly by a sign
change in Eq.~(\ref{eq:derivationAloha}); moreover, the case of
non-saturated arrivals $A_1(t)$ follows similarly by plugging-in
their moment generating function (see~\cite{CiucuISIT11}).

Next we give a general result for computing the upper and lower
bounds on the end-to-end throughput capacity for a flow crossing
$k$ hops.

\begin{theorem}{({\sc{Capacity Bounds (Upper and Lower Bounds) - Aloha}})}\label{th:e2ecap}
Consider a flow crossing $k$ hops. Assume that the
impulse-response process $S_j(s,t)$, at each hop $j$, satisfies
the following bounds on the MGF and Laplace transforms:
$E\left[e^{\theta S_j(s,t)}\right]\leq e^{\theta
r_j(\theta)(t-s)}$ and $E\left[e^{-\theta S_j(s,t)}\right]\leq
e^{-\theta r_j(-\theta)(t-s)}$ for all $\theta>0$. Let
$r(\theta)=\max_j r_j(\theta)$ and $r(-\theta)=\min_j
r_j(-\theta)$. Assume also that $S_j(s,t)$ are statistically
independent over non-overlapping intervals. Then, for some
$\eps>0$, a probabilistic lower bound on the capacity rate is for
all $t\geq0$
\begin{equation}
\lambda_t^L=\sup_{\theta>0}\left\{r(-\theta)+\frac{\log{\eps}-\log{\binom{t+k-1}{k-1}}}{\theta
t}\right\}~,\label{eq:lbAloha}
\end{equation}
The corresponding upper bound is
\begin{equation}
\lambda^U_t=\inf_{\theta>0}\left\{r(\theta)-\frac{\log{\eps}}{\theta
t}\right\}~.\label{eq:ubAloha}
\end{equation}
\end{theorem}

{\sc Proof.}~Denote $A(t)$ and $D(t)$ the arrival and departure
processes of the flow at the first node and last node,
respectively; assume the saturated condition that $A(t)=\infty$.
Applying the end-to-end service curve from Eq.~(\ref{eq:e2econv}),
extended to $k$ hops, we can write
\begin{eqnarray*}
\P\left(D(t)\leq \lambda_t^L t\right)&\leq&\P\left(A\conv S(t)\leq
\lambda_t^L t\right)\\
&=&\P\left(S_1\conv S_2\conv\dots\conv S(t)\leq\lambda_t^L
t\right)~.
\end{eqnarray*}
Letting $u_0=0$ and $u_k=t$ we can continue bound the probability
using the Chernoff bound:\begin{eqnarray*}
&&\hspace{-1cm}\P\left(\inf_{0\leq u_1\leq\dots\leq u_{k-1}\leq
t}\sum_{j=1}^k
S_j\left(u_{j-1},u_j\right)\leq \lambda_t^L t\right)\notag\\
&&\leq \sum_{u_0\leq\dots\leq u_{k}}\prod_{j=1}^k E\left[^{-\theta
S_{j}(u_{j-1},u_j)}\right]e^{\theta \lambda_t^L t}\\
&&\leq \sum_{0\leq u_1\leq\dots\leq u_{k}}\prod_{j=1}^k e^{-\theta r_j(-\theta)(u_j-u_{j-1})}e^{\theta\lambda_t^L t}\\
&&=\binom{t+k-1}{k-1}e^{-\theta\left(r(-\theta)-\lambda_t^L\right)t}~,
\end{eqnarray*}
where the binomial term is the number of combinations with
repetition. Equalizing the last term to $\eps$ gives the bound
$\lambda_t^L$.

In turn, for the upper bound, we can write
\begin{eqnarray*}
&&\hspace{-1cm}\P\left(D(t)\geq \lambda^U_t
t\right)=\P\left(A\conv S(t)\geq
\lambda^U_t t\right)\notag\\
&&=\P\left(S_1\conv S_2\conv\dots\conv S(t)\geq\lambda^U_t
t\right)\notag\\
&&\leq\inf_{u_1\leq\dots\leq u_k}\P\left(\sum_{j=1}^k
S_j(u_{j-1},u_j)\geq\lambda^U_t t\right)\\
&&\leq \inf_{u_1\leq\dots\leq u_k}\prod_{j=1}^k E\left[^{\theta
S_{j}(u_{j-1},u_j)}\right]e^{-\theta \lambda_t^U t}\\
&&\leq e^{-\theta\left(\lambda_t^U-r(\theta)\right)t}~.
\end{eqnarray*}
Equalizing the obtained bound with $\eps$ and solving for
$\lambda^U_t$ we obtain the value from Eq.~(\ref{eq:ubAloha}),
which completes the proof.~\hfill $\Box$


\subsection{CSMA/CA}\label{sec:csmaca}

The CSMA/CA protocol was motivated by the need to increase the
(very) low capacity of Aloha, while preserving the distributive
aspect of the protocol. One key idea is to prevent collisions from
happening by enabling nodes to `listen to the channel' before
transmitting. The other key idea is that once a node perceives the
channel as being busy, the node enters in an exponentially
distributed backoff mode.

\begin{figure}[t]
\centerline{\epsfig{figure=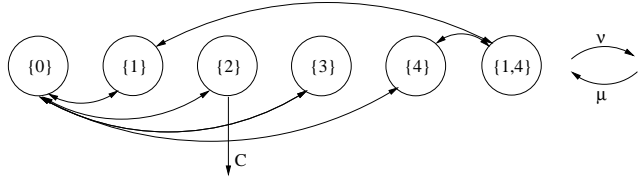,width=3.3in,keepaspectratio}
} \caption{A Markov Modulated Transmission Process (MMTP) for link
$[A_2\rightarrow A_3]$} \label{fig:csma}
\end{figure}

We use a simplified CSMA/CA protocol, developed by Durvy et
al.~\cite{Durvy09}, which retains the key features of CSMA/CA. For
the network from Figure~\ref{fig:5Nodes}.(a-b), the construction
of the MMTP processes, and also of the impulse-response processes
$S_i(s,t)$, follows similarly as for centralized scheduling and
Aloha. For the link $[A_2\rightarrow A_3]$, the underlying MMTP,
and also the modulating (now continuous-time) Markov process
$X(t)$, are depicted in Figure~\ref{fig:csma}. $X(t)$ is
constructed exactly as in~\cite{Durvy09}, where $\nu^{-1}$ and
$\mu^{-1}$ denote the average backoff and transmission times. The
interpretation of the states is identical as for centralized
scheduling (see Figure~\ref{fig:centralAloha}.(a)); while $X(t)$
delves in the new state $\{0\}$, all nodes are in a backoff mode.
Ignoring the details of switching from discrete to continuous
time, the MMTP is defined as for Aloha (see
Eq.~(\ref{eq:mmtpAloha})), i.e.,

\begin{equation*}
C_{X(t)}=\left\{\begin{array}{lll}C&,&\textrm{if}~X(t)=\textrm{\{2\}}\\
0&,&\textrm{otherwise~,}\end{array}\right.
\end{equation*}
and the impulse-response $S_2(s,t)$ is defined as in
Eq.~(\ref{eq:smmtp}). The MMTPs for the other links are defined
similarly (e.g., for $[A_1\rightarrow A_2]$, the only change is
that $C_{X(t)}=C$ when $X(t)\in\left\{\{1\},\{1,4\}\right\}$).

Unlike Aloha, CSMA/CA is subject to more compounded calculations
of the e2e throughput capacity. Note that the first two lines from
Eq.~(\ref{eq:derivationAloha}) still hold. The last line, however,
does not hold anymore since $S_1(u,s)$ and $S_2(s,t)$ are
\textit{not} statistically independent, even over non-overlapping
intervals. The reason lies in the Markov modulating process $X(t)$
which does \textit{not} have independent increments (as in
Eq.~(\ref{eq:indincr}) for Aloha). The immediate solution is to
apply H\"{o}lder's inequality to bound $E\left[e^{-\theta
S_1(s)}e^{-\theta S_2(s,t)}\right]$, as suggested for the Aloha
case in extended (not-necessarily line) networks.

The last unaddressed issue concerns the computation of the Laplace
transforms for the impulse-responses $S_i(s,t)$'s. Since these
processes are Markov arrival processes (MAPs), their Laplace
transforms can be computed using standard techniques from
teletraffic theory (see Courcoubetis and
Weber~\cite{Courcoubetis96}). Let us briefly reproduce such
techniques and compute in particular
\begin{equation*}L_{2,t}:=E\left[e^{-\theta S_2(t)}\right]~,
\end{equation*} for some $\theta>0$. Denote the six states of the MAP from
Figure~\ref{fig:csma} by the numbers $1,2,\dots,6$, and the
elements of the generator matrix $P$ by $p_{i,j}$, i.e.,
\begin{equation*}
P=\left(\begin{array}{cccccc}-4\nu&\nu&\nu&\nu&\nu&0\\
\mu&-(\mu+\nu)&0&0&0&\nu\\
\mu&0&-\mu&0&0&0\\
\mu&0&0&-\mu&0&0\\
\mu&0&0&0&-(\mu+\nu)&\nu\\
0&\mu&0&0&\mu&-2\mu
\end{array}\right)~.
\end{equation*}

Denote also the conditional Laplace transforms
$L_{2,i,t}:=E\left[e^{-\theta S_2(t)}\mid X(0)=i\right]$, i.e.,
conditioned on the initial state of the Markov chain $X(t)$, which
is assumed to start in steady-state. For some initial state (e.g.,
$i=3$) we have the backward equation
\begin{eqnarray*}
L_{2,3,t+\Delta t}&=&E\left[e^{-\theta S_2(\Delta t)}\mid X(0)=3\right]\\
&&\qquad\sum_jE\left[e^{-\theta S_2(\Delta t,t+\Delta
t)}\mid X(\Delta t)=j\right]p_{3,j}\\
&=&e^{-\theta C\Delta t}\left(L_{2,0,t}\mu\Delta
t+L_{2,3,t}\left(1-\mu\Delta t\right)+o(\Delta t)\right)~,
\end{eqnarray*}
where $\lim_{\Delta t\rightarrow 0}\frac{o(\Delta t)}{\Delta
t}=0$. In the last line we used the stationarity of $S_2(t)$.
Using the Taylor's expansion $e^{-\theta C\Delta t}=1-\theta
C\Delta t+o(\Delta t)$, rearranging terms, and taking the limit
$\Delta t\rightarrow 0$ it follows that
\begin{equation}
\frac{\partial L_{2,3,t}}{\partial
t}=L_{2,0,t}\mu-L_{2,3,t}(\mu+\theta C)~.\label{eq:pde2}
\end{equation}
One can proceed similarly to derive the PDE's of the other
$L_{2,i,t}$'s for $i\neq 3$, and arrive at the system of PDE's
\begin{eqnarray}
\frac{\partial \mbox{\boldmath${L_{2,t}}$}}{\partial
\mbox{\boldmath${t}$}}
=B_2\mbox{\boldmath${L_{2,t}}$}~,\label{eq:pdev}
\end{eqnarray}
where $\mbox{\boldmath${L_{2,t}}$}=(L_{2,1,t},\dots,L_{2,6,t})^T$
and $B_2$ is a matrix whose lines are formed according to
Eq.~(\ref{eq:pde2}), i.e.,
\begin{equation*}
B_2=\left(\begin{array}{cccccc}-4\nu&\nu&\nu&\nu&\nu&0\\
\mu&-(\mu+\nu)&0&0&0&\nu\\
\mu&0&-(\mu+\theta C)&0&0&0\\
\mu&0&0&-\mu&0&0\\
\mu&0&0&0&-(\mu+\nu)&\nu\\
0&\mu&0&0&\mu&-2\mu
\end{array}\right)~.
\end{equation*}
Note that $B_2$ differ from the generator matrix $P$ by the term
$-\mu C$ on the position $(3,3)$, which is due to the transmission
at rate $C$ while in state $3$ (i.e., the state labelled `$\{2\}$'
from Figure~\ref{fig:csma}). Solving for Eq.~(\ref{eq:pdev}) we
get the solution
\begin{equation*}
\mbox{\boldmath${L_{2,t}}$}=\sum_{i=1}^6 c_ie^{\lambda_i
t}\mbox{\boldmath${x_i}$}~,
\end{equation*}
where $\lambda_i$ and $\mbox{\boldmath${x_i}$}$ are the
eigenvalues and eigenvectors, respectively, of the matrix $B_2$.
The coefficients
$\mbox{\boldmath${c}$}=\left(c_1,\dots,c_n\right)^T$ can be
determined from the initial condition
$\mbox{\boldmath${c}$}=X^{-1}\mbox{\boldmath${1}$}$,
$X=\left(\mbox{\boldmath${x_1}$},\dots,\mbox{\boldmath${x_n}$}\right)$
and $\mbox{\boldmath${1}$}=\left(1,\dots,1\right)^T$. Because the
modulating process starts in the steady-state, we obtain the
solution of the Laplace transform $L_{2,t}=\sum_{i=1}^{6}\pi_i
L_{2,i,t}$. This takes the hyperexponential form
\begin{equation}
L_{2,t}=E\left[e^{-\theta S_2(t)}\right]=\sum_{i=1}^6
L_{2,i}e^{\lambda_i t}~,\label{eq:mgfcond0}
\end{equation}
where $L_{2,i}=c_i\sum_j \pi_jX_{j,i}$. Since $\sum_i L_{2,i}=1$,
we have the simplified bound
\begin{equation}
E\left[e^{-\theta S_2(t)}\right]\leq e^{\lambda_6
t}~,\label{eq:spectrad}
\end{equation}
where, by convention, $\lambda_6=\max_i\lambda_i$ is the spectral
radius of the matrix $B_2$.

The other Laplace transforms can be computed similarly, by
suitably modifying the MMTP as mentioned earlier. The only
difference, when computing $E\left[e^{-\theta S_j(t)}\right]$, is
that the corresponding matrix $B_j$ is given by
\begin{equation}
B_j=P-\theta CI_j\label{eq:bj}
\end{equation}
where $I_{j,i,i}=1$ if there is a transmission at rate $C$ in the
corresponding MMTP, and $I_{j,i,l}=0$ otherwise (as an example,
when $j=4$, then $I_{j,5,5}=1$, $I_{j,6,6}=1$, and $I_{j,i,l}=0$
otherwise).

Compared to Aloha, the drawback of the closed-form results
obtained for CSMA/CA is that they depend on the
eigenvalues/eigenvectors of the matrix $B$ (appearing in the
solution of Eq.~(\ref{eq:pdev})), and are thus not easily amenable
to convex optimizations. The next theorem extends
Theorem~\ref{th:e2ecap} to the CSMA/CA case.

\medskip


\begin{theorem}{({\sc{Capacity Bounds (Upper and Lower Bounds)~-~CSMA/CA}})}\label{th:e2ecapcsma}
Consider a flow crossing $k$ hops and the corresponding MMTP being
modulated by a generator matrix $P$. For $j=1,2,\dots,k$, let
$\lambda_j$ as the spectral radiuses of the matrixes $B_j$
constructed as in Eq.~(\ref{eq:bj}) with $\theta$ replaced by
$k\theta$.  Let $r_j(-k\theta):=\frac{\lambda_j}{-k\theta}$ and
$r(-k\theta)=\min_j r_j(-k\theta)$. Then, for some $\eps>0$, a
probabilistic lower bound on the capacity rate is for all $t\geq0$
\begin{equation}
\lambda_t^L=\sup_{\theta>0}\left\{r(-k\theta)+\frac{\log{\eps}-\log{\binom{t+k-1}{k-1}}}{\theta
t}\right\}~,\label{eq:lbcsma}
\end{equation}
In turn, for the upper bound, let $\lambda_j$ as the spectral
radiuses of the matrixes $B_j$ constructed as in Eq.~(\ref{eq:bj})
with $\theta$ replaced by $-k\theta$, for $j=1,2,\dots,k$. Let
also $r_j(k\theta):=\frac{\lambda_j}{k\theta}$ and
$r(k\theta)=\max_j r_j(k\theta)$. Then, for some $\eps>0$, a
probabilistic upper bound on the capacity rate is for all $t\geq0$
\begin{equation}
\lambda^U_t=\inf_{\theta>0}\left\{r(k\theta)-\frac{\log{\eps}}{\theta
t}\right\}~.\label{eq:ubcsma}
\end{equation}
\end{theorem}

{\sc Proof.}~The proof proceed similarly as the proof of
Theorem~\ref{th:e2ecap}, with the observation that one has to
account for the fact that the impulse-responses $S_j(s,t)$ are not
anymore statistically independent (even over non-overlapping
intervals).

The first step is identical, i.e.,
\begin{eqnarray*}
\P\left(D(t)\leq \lambda_t^L t\right)&\leq&\P\left(A\conv S(t)\leq
\lambda_t^L t\right)\\
&=&\P\left(S_1\conv S_2\conv\dots\conv S(t)\leq\lambda_t^L
t\right)~.
\end{eqnarray*}
For the second step, we additionally rely on H\"{o}lder's
inequality:\begin{eqnarray*} &&\hspace{-1cm}\P\left(\inf_{0\leq
u_1\leq\dots\leq u_{k-1}\leq t}\sum_{j=1}^k
S_j\left(u_{j-1},u_j\right)\leq \lambda_t^L t\right)\notag\\
&&\leq \sum_{u_0\leq\dots\leq u_{k}}\prod_{j=1}^k
\left(E\left[^{-k\theta
S_{j}(u_{j-1},u_j)}\right]\right)^{\frac{1}{k}}e^{\theta \lambda_t^L t}\\
&&\leq \sum_{0\leq u_1\leq\dots\leq u_{k}}\prod_{j=1}^k e^{-\theta r_j(-k\theta)(u_j-u_{j-1})}e^{\theta\lambda_t^L t}\\
&&=\binom{t+k-1}{k-1}e^{-\theta\left(r(-k\theta)-\lambda_t^L\right)t}~,
\end{eqnarray*}
where the binomial term is the number of combinations with
repetition. Equalizing the last term to $\eps$ gives the bound
$\lambda_t^L$.

In turn, for the upper bound, we can write
\begin{eqnarray*}
&&\hspace{-1cm}\P\left(D(t)\geq \lambda^U_t
t\right)=\P\left(A\conv S(t)\geq
\lambda^U_t t\right)\notag\\
&&=\P\left(S_1\conv S_2\conv\dots\conv S(t)\geq\lambda^U_t
t\right)\notag\\
&&\leq\inf_{u_1\leq\dots\leq u_k}\P\left(\sum_{j=1}^k
S_j(u_{j-1},u_j)\geq\lambda^U_t t\right)\\
&&\leq \inf_{u_1\leq\dots\leq u_k}\prod_{j=1}^k
\left(E\left[^{k\theta
S_{j}(u_{j-1},u_j)}\right]\right)^{\frac{1}{k}}e^{-\theta \lambda_t^U t}\\
&&\leq e^{-\theta\left(\lambda_t^U-r(k\theta)\right)t}~.
\end{eqnarray*}
Equalizing the obtained bound with $\eps$ and solving for
$\lambda^U_t$ we obtain the value from Eq.~(\ref{eq:ubcsma}),
which completes the proof.~\hfill $\Box$

En passant, we point out that existing multi-hop results
(especially for CSMA MACs) rely on independence assumptions across
hops (e.g,~\cite{GaoCL06,XieH09}), in addition to the saturation
assumption at the relay nodes. In fact, even the seminal
single-hop result obtained by Bianchi~\cite{Bianchi00} relies on
the artificial assumption that nodes independently see the system
in the steady-state. We raise the awareness that
\textit{convenient} statistical assumptions are inherently prone
to incorrect results, even when given as asymptotic scaling laws,
as long as independence assumptions extend over a number of hops
as a function of the total number of hops. This pitfall has been
also pointed out even in an M/M/1 packet tandem network
(\textit{wired}), subject to the classical Kleinrock's
independence assumption (see~\cite{BuLiCi11}).

\section{Application: Single-Hop vs. Multi-Hop}\label{sec:shmh}
\begin{figure}[t]
\vspace{-.15cm}
\begin{center}
\hspace{-0.1cm}\shortstack{
\includegraphics[width=0.4\linewidth,keepaspectratio]{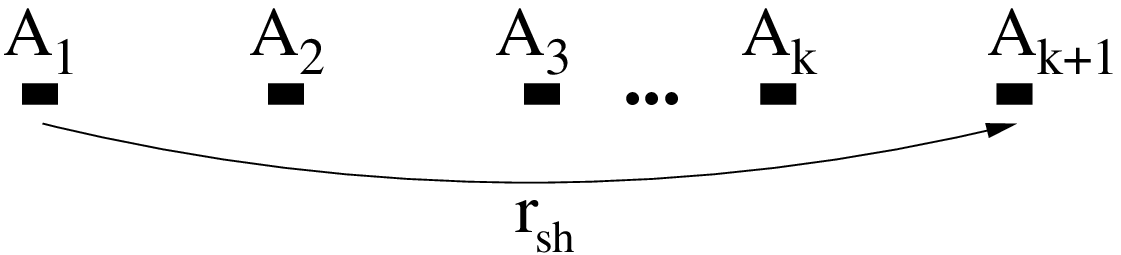}
\\
{\footnotesize (a) Single-hop} } \hspace{-0cm}\shortstack{
\includegraphics[width=0.4\linewidth,keepaspectratio]{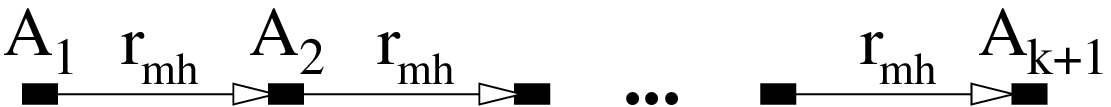}
\\
{\footnotesize (b) Multi-hop} } \caption{Which strategy should
node $A_1$ choose in order to transmit to $A_{k+1}$? (all nodes
hear each other, all are saturated, and
$r_{\textrm{sh}}<r_{\textrm{mh}}$)} \label{fig:shvsmh}
\end{center}
\vspace{-0.25cm}
\end{figure}

In this section we give an illustrative example on using the
\textit{finite time and space} key features of our capacity bounds
for the following problem. Consider the network from
Figure~\ref{fig:shvsmh} with $k+1$ nodes, all in the interference
range of each other, and all being saturated and attempting to
access the channel using either the Aloha or CSMA/CA protocols.
Given that node $A_1$ intends to transmit to node $A_{k+1}$, the
problem concerns choosing between the following two routing
strategies:

\begin{enumerate}
\item{Single-hop}: Node $A_1$ directly transmits to node $A_{k+1}$
at rate $r_{\textrm{sh}}$.

\item{Multi-hop}: Node $A_1$ transmits using the nodes
$A_2,~A_3,\dots,A_k$ as relays; the rate for each transmission
$\left[A_j\rightarrow A_{j+1}\right]$ is $r_{\textrm{mh}}$.
\end{enumerate}
In order to avoid a trivial answer we assume that
$r_{\textrm{sh}}<r_{\textrm{mh}}$.

\begin{figure}[h]
\begin{center}
\hspace{-0cm}\shortstack{
\includegraphics[width=0.4\linewidth,keepaspectratio]{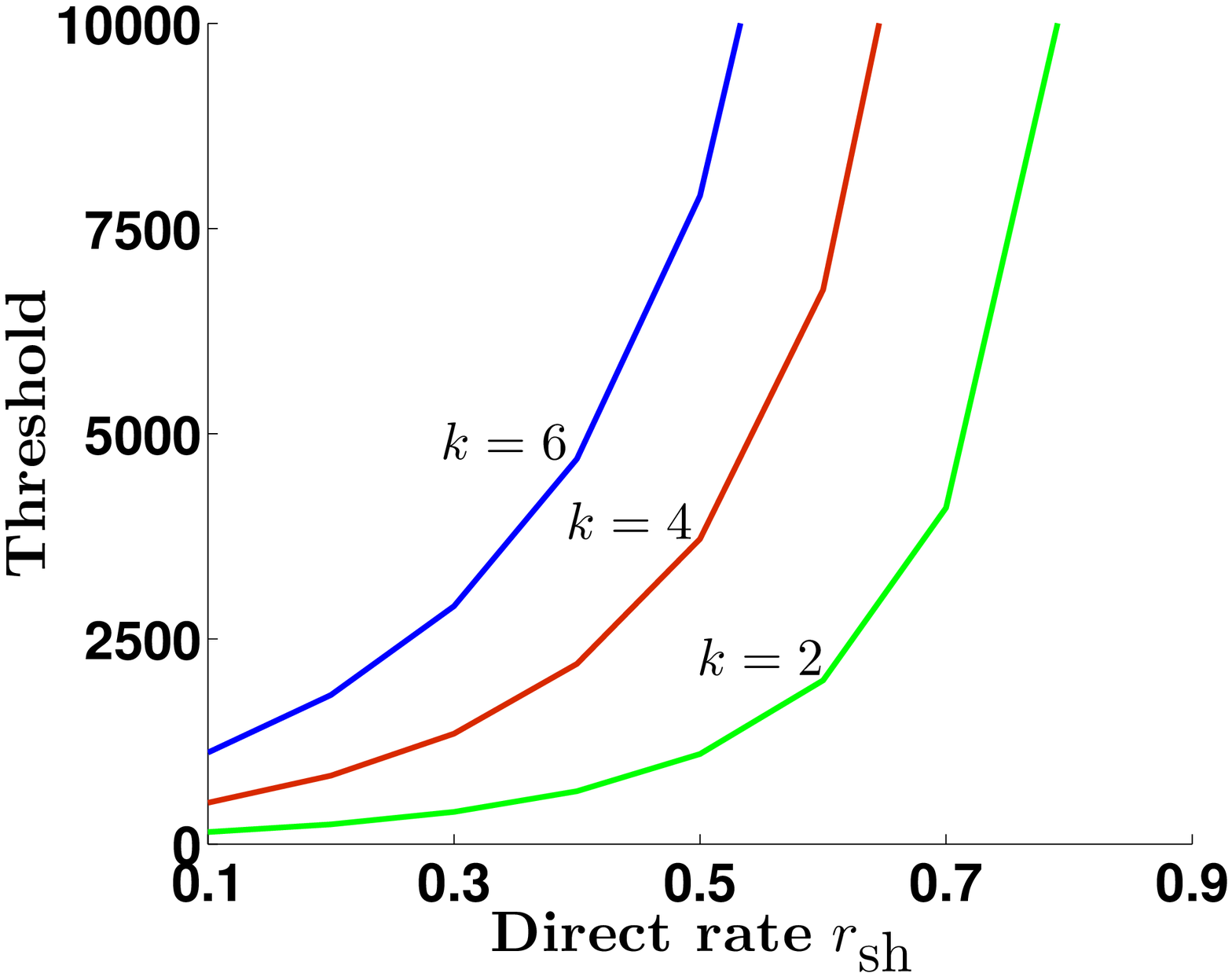}
\\
{\footnotesize (a) Aloha} } \shortstack{
\includegraphics[width=0.4\linewidth,keepaspectratio]{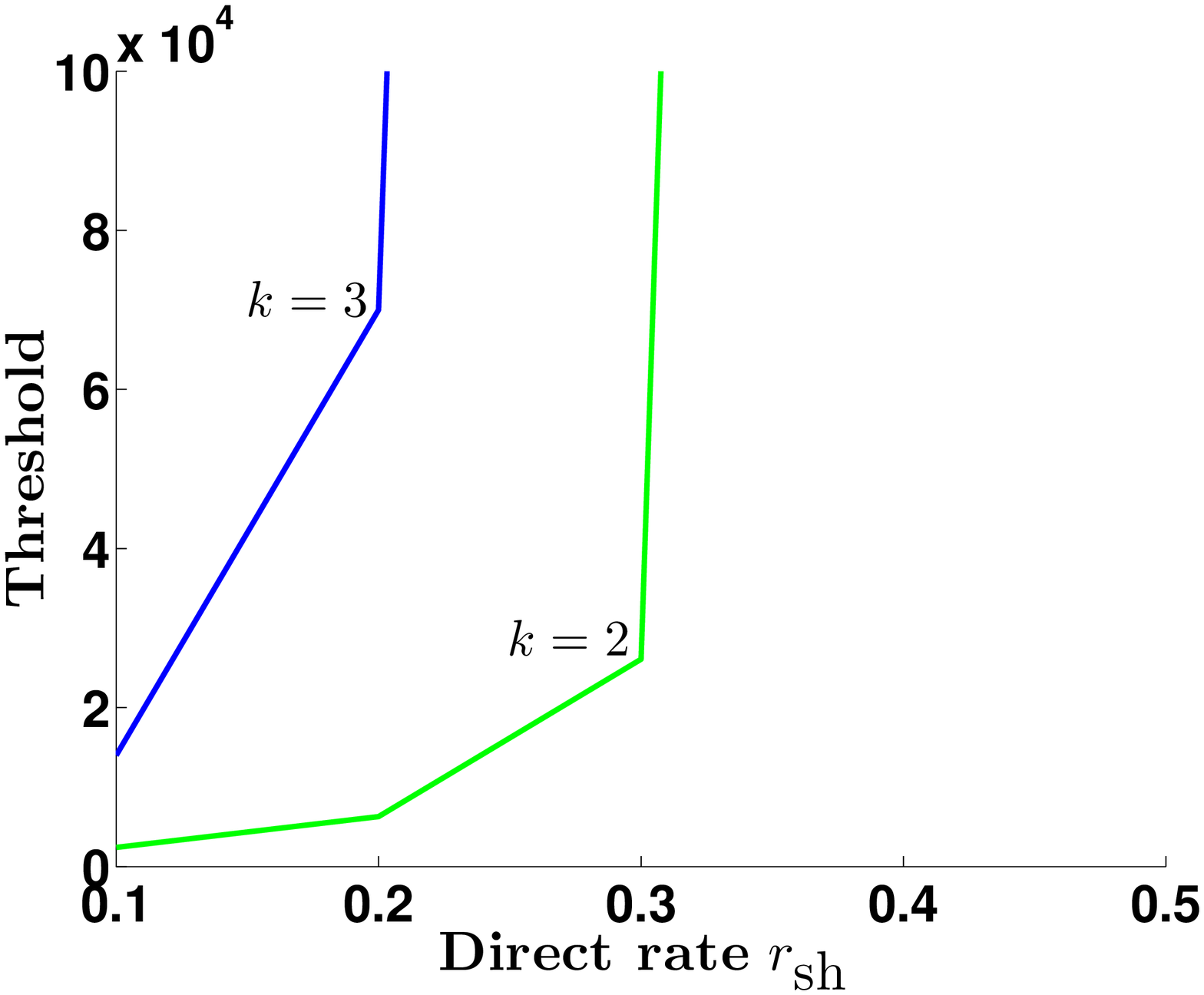}
\\
{\footnotesize (b) CSMA/CA} } \caption{The time scale (threshold)
after which the multi-hop is more advantageous than the single-hop
strategy ($\eps=10^{-3}$, normalized per-hop rate
$r_{\textrm{mh}}=1$, variable direct rate $r_{\textrm{sh}}$, Aloha
transmission probability $p=\frac{1}{k}$, and $\nu=\mu=0.1$ for
CSMA/CA)} \label{fig:threshold}
\end{center}
\vspace{-0.25cm}
\end{figure}

Figure~\ref{fig:threshold} illustrates the threshold at which the
multi-hop strategy is more advantageous. More concretely, the
values displayed (i.e., the `Threshold') are the time scales at
which the \textit{lower-bound} for the multi-hop transmission is
larger than the \textit{upper-bound} for the single-hop
transmission\footnote{The lower and upper bounds are applications
of Theorems~\ref{th:e2ecap} and~\ref{th:e2ecapcsma}.}. Both (a)
and (b) indicate the intuitive facts that the `Threshold' is
exponential in the relative direct rate $r_{\textrm{sh}}$ and also
increasing in the number of hops $k$. In (b), for CSMA/CA, the
benefits of multi-hop routing hold only for very low relative
direct rate $r_{\textrm{sh}}$ and quickly vanish by increasing
$k$. We point out however that this quick blow-up may be due to
the loose underlying upper bounds on the CSMA/CA per-flow
capacity. In the case of Aloha, however, both the lower and upper
bounds are reasonably tight. For actual illustrations of the
numerical tightness of the bounds we refer to
Section~\ref{sec:numerics}.

The above routing problem has been debated in different settings
such as wireless mesh and sensor networks. Experimental results by
De Couto~\et~\cite{Couto03} showed that minimizing the hop count
is not always the best option as long hops may incur a high packet
error rate. Jain~\et~\cite{jain03} showed that, due to
interference, shortest paths with long hops may not provide the
best performance. In contrast, there are several results
supporting long-hop routing. Haenggi and
Puccinelli~\cite{Haenggi05} provided many reasons why short-hop
routing is not as beneficial as it seems to be. Moreover, in
energy limited networks such as sensor networks, long-hop routing
may also be preferable (see Ephremides~\cite{Ephremides02} and
Bj\"{o}rnemo \et~\cite{Bjornemo07}).

Our contribution to this debate is to bring a new perspective on
single vs. multi-hop routing by focusing on the underlying time
scale. Concretely, we provided theoretical evidence that multi-hop
routing is more advantageous in the long-run for Aloha. In turn,
in the case of CSMA/CA, the advantage of multi-hop vanishes in
most cases. We raise however the awareness that, for the purpose
of analytical tractability, our results are restricted to a line
network and no frequency or power management being accounted for.

\section{Numerical Results}\label{sec:numerics} Here we briefly illustrate the
numerical tightness of the derived lower and upper bounds on the
\textit{transient throughput rate} from Theorems~\ref{th:e2ecap}
and~\ref{th:e2ecapcsma}. We consider the network from
Figure~\ref{fig:shvsmh}.(b) in which node $A_1$ transmits to node
$A_{k+1}$ in a multi-hop fashion. We use the parameters per-hop
rate $r_{\textrm{mh}}=1$, $p=\frac{1}{k}$ for Aloha,
$\frac{1}{\nu}=\frac{1}{\mu}=10$ for CSMA/CA, and a violation
probability $\eps=10^{-3}$.

\begin{figure}[h]
\begin{center}
\shortstack{
\includegraphics[width=0.45\linewidth,keepaspectratio]{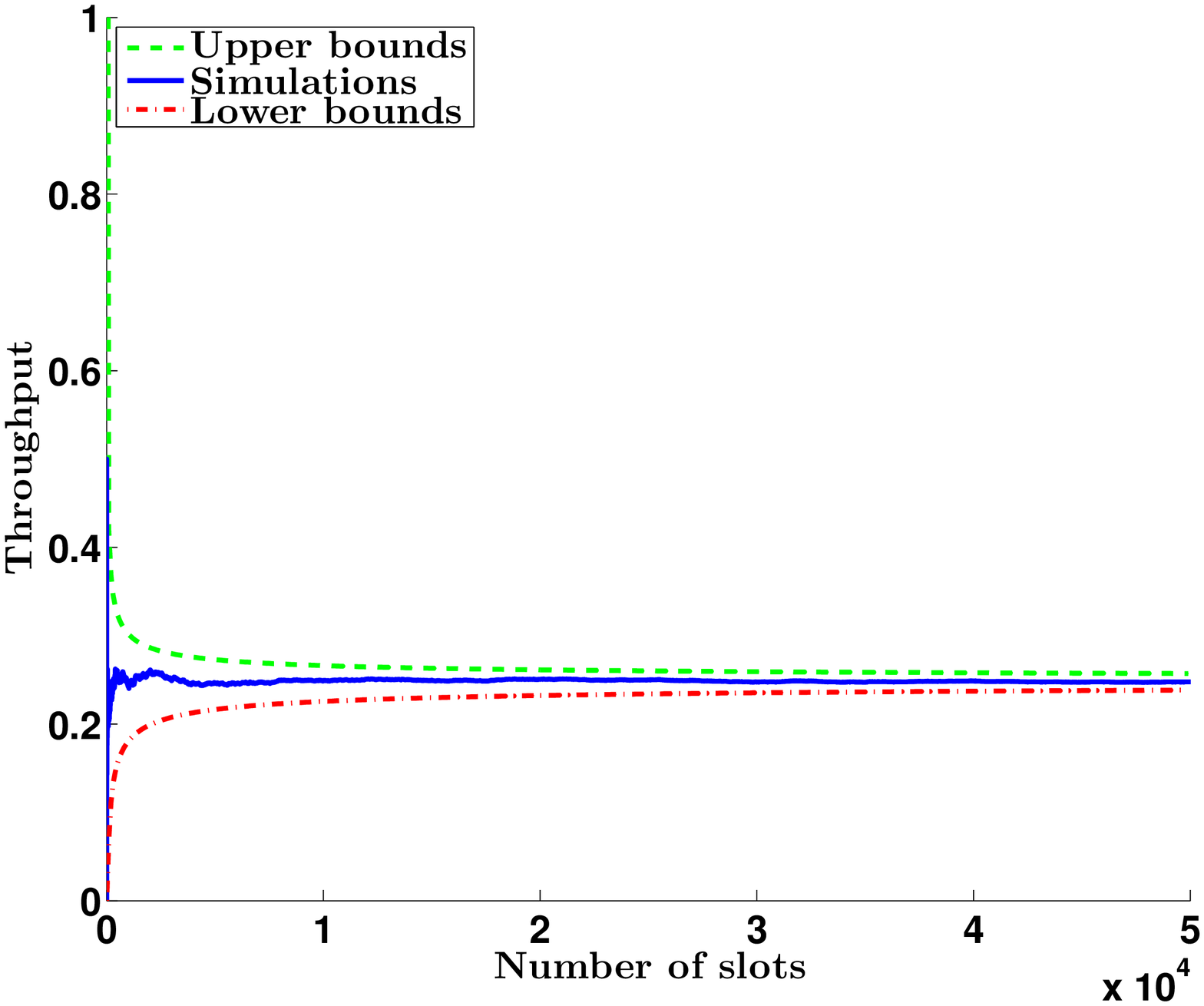}
\\
{\footnotesize (a) Aloha, $k=2$} } \shortstack{
\includegraphics[width=0.45\linewidth,keepaspectratio]{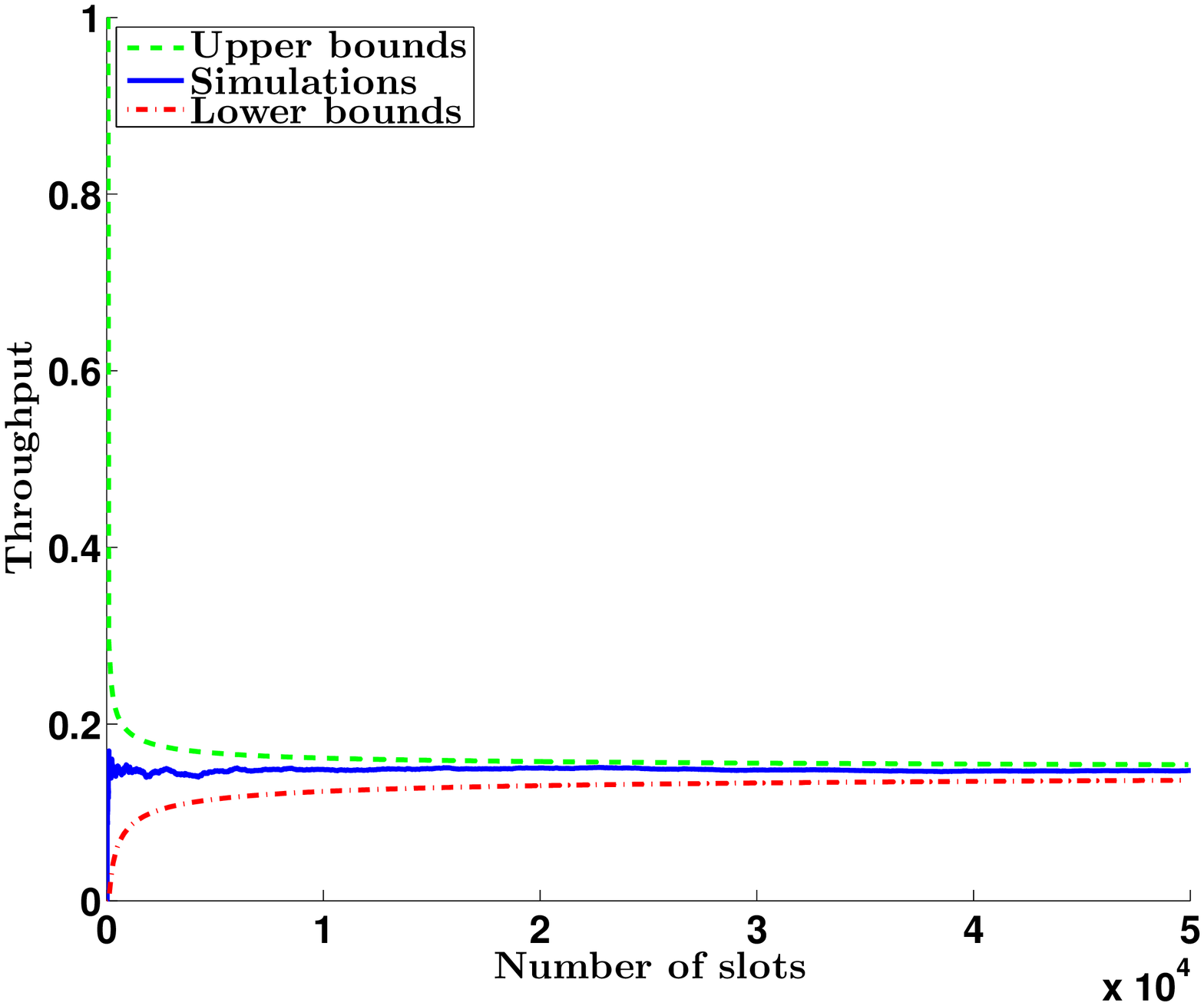}
\\
{\footnotesize (b) Aloha, $k=3$} } \shortstack{\hspace{0cm}
\includegraphics[width=0.45\linewidth,keepaspectratio]{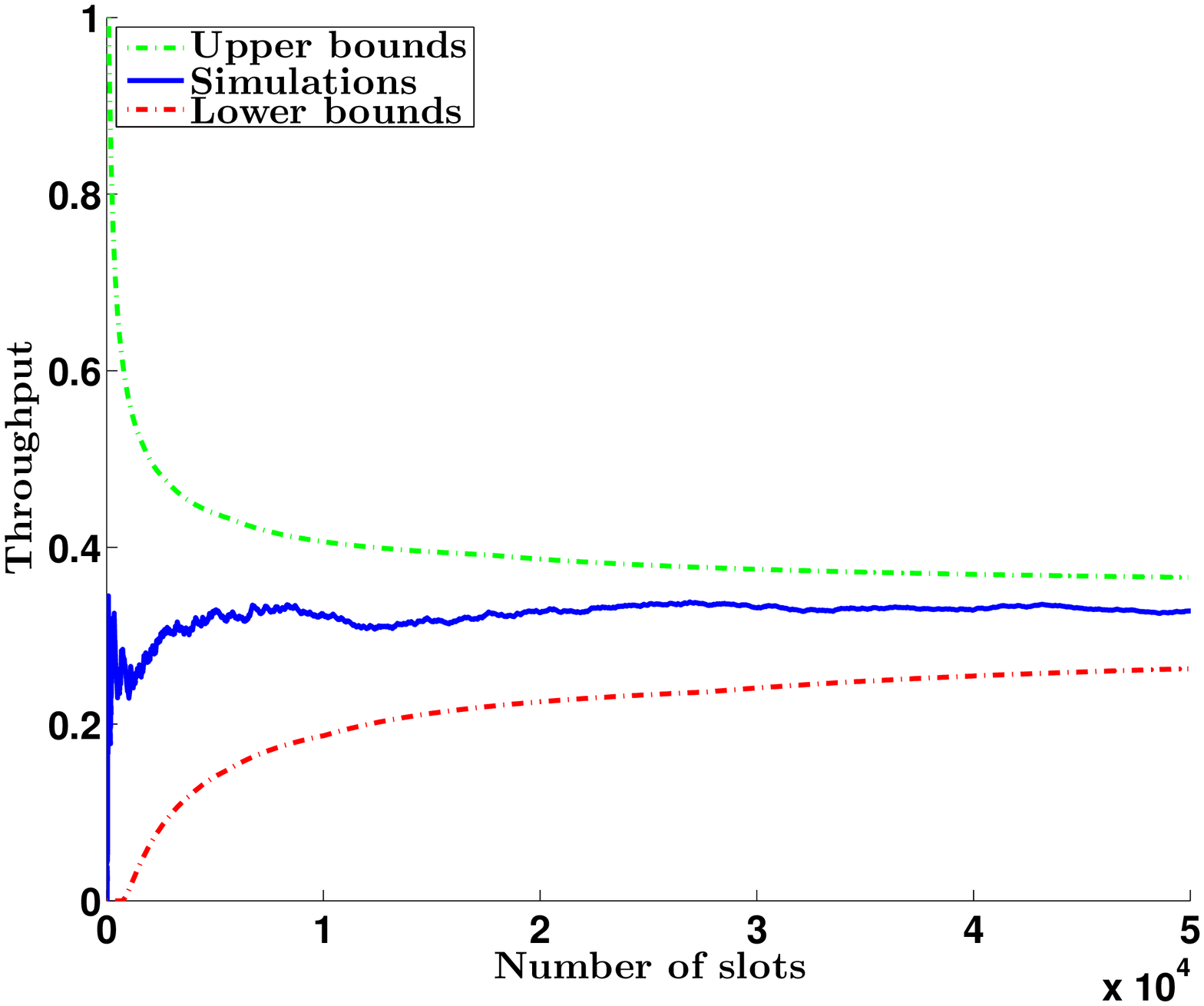}
\\
{\footnotesize (c) CSMA/CA, $k=2$} } \shortstack{
\includegraphics[width=0.45\linewidth,keepaspectratio]{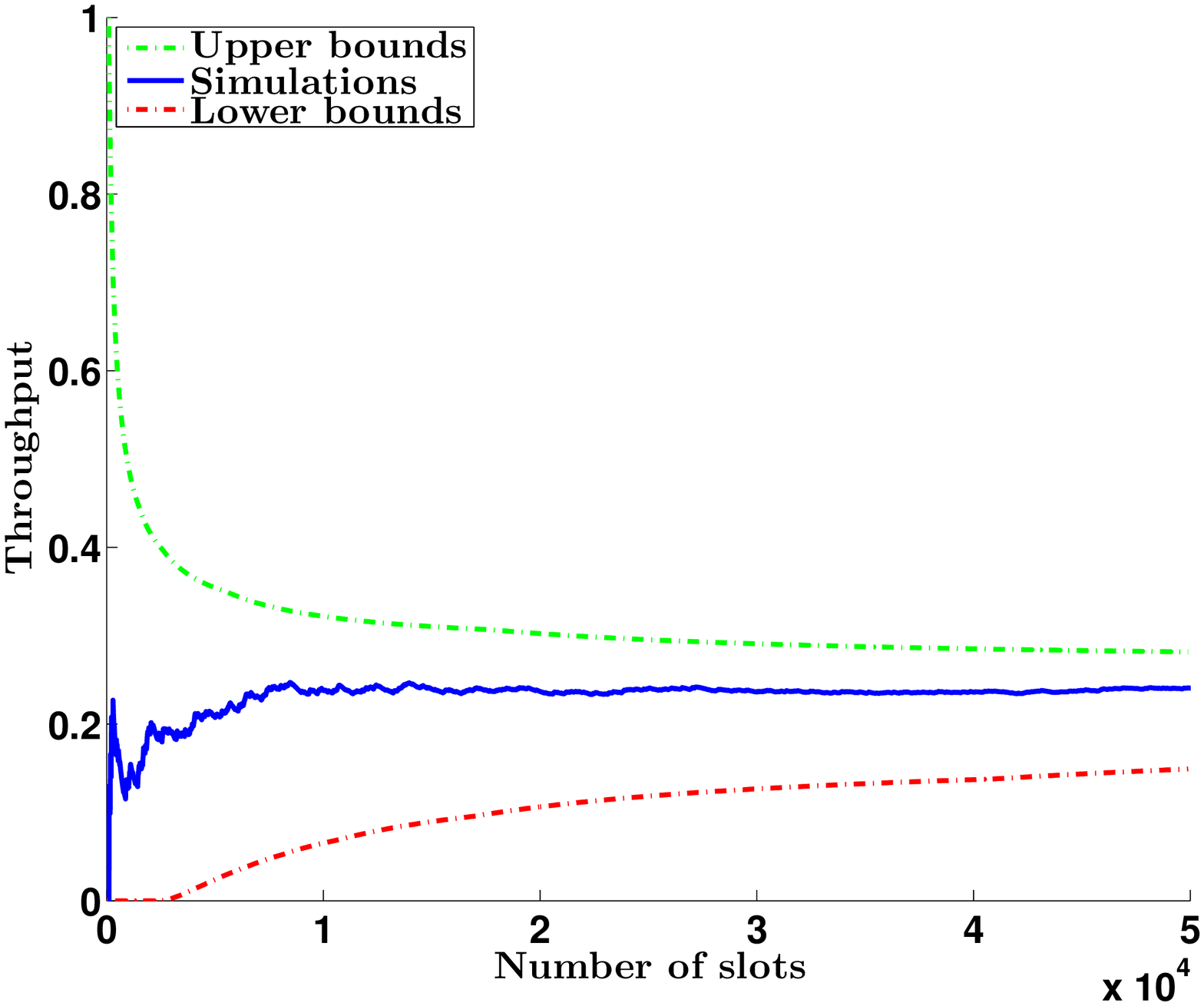}
\\
{\footnotesize (d) CSMA/CA, $k=3$} } \caption{Throughput rates as
a function of the number of time slots.} \label{fig:numerics23}
\end{center}
\end{figure}

Figure~\ref{fig:numerics23} indicates that the upper and lower
bounds for Aloha are quite tight. For CSMA/CA, however, only the
upper bounds remain reasonably tight whereas the lower bounds tend
to degrade with the number of hops $k$. This is due to the
underlying application of the Boole's inequality, which is known
to be loose in the case of correlated arrivals (see,
e.g.,~\cite{CiPoSc13}).

\section{Conclusions}\label{sec:discussion}
We have presented the key ingredients of a \textit{unified}
system-theoretic methodology to compute the per-flow capacity in
finite time and space network scenarios, and for three MAC
protocols: centralized scheduling, Aloha, and CSMA/CA. We have
also confirmed the anecdotal practical conservative nature of
alternative asymptotic results, by closely analyzing a widely used
double-limit argument. Moreover, we have demonstrated that our
finite time/space results can lend themselves to engineering
insight, i.e., on the time scales at which multi-hop routing
becomes more advantageous than single-hop routing.

The presented methodology faced however the following dilemma
concerning analytical tractability vs. the level of modelling
details. One one hand, we have managed to employ a rigorous
mathematical analysis. On the other hand, due to the hardness of
dealing with spatio-temporal correlations in non-Jackson queueing
networks, we have restricted to a line network and simplified MAC
protocols, while ignoring physical layer considerations. Moreover,
we have employed bounding techniques from the effective bandwidth
literature, and whose numerical accuracy is problematic in the
case of correlated processes (e.g.,
Choudhury~\et~\cite{choudhury96squeezing}). While such techniques
produced good estimates for the Aloha case, there is a need for
advanced techniques to properly account for the underlying
correlations in CSMA/CA. Nevertheless we believe that the
advocated system-theoretic approach has the potential to
contribute to the development of the long desirable
\textit{functional network information theory} (see
Andrews~\et~\cite{Andrews08commag}).

%
\bibliographystyle{abbrv}
\bibliography{../../stat}  

\newcommand{\noopsort}[1]{}\providecommand{\noopsort}[1]{}
\begin{thebibliography}{10}

\bibitem{Abramson70}
N.~Abramson.
\newblock The {A}loha system: another alternative for computer communications.
\newblock In {\em Proceedings of {AFIPS} Joint Computer Conferences}, pages
  281--285, 1970.

\bibitem{Andrews08commag}
J.~G. Andrews, N.~Jindal, M.~Haenggi, R.~Berry, S.~Jafar, D.~Guo,
  S.~Shakkottai, R.~Heath, M.~Neely, S.~Weber, and A.~Yener.
\newblock Rethinking {Information Theory} for mobile ad hoc networks.
\newblock {\em IEEE Communications Magazine}, 46(12):94--101, Dec. 2008.

\bibitem{Bianchi00}
G.~Bianchi.
\newblock Performance analysis of the {IEEE} 802.11 distributed coordination
  function.
\newblock {\em IEEE Journal on Selected Areas in Communications},
  18(3):535--547, Mar. 2000.

\bibitem{Bjornemo07}
E.~Bj\"{o}rnemo, M.~Johansson, and A.~Ahl\'{e}n.
\newblock Two hops is one too many in an energy-limited wireless sensor
  network.
\newblock In {\em IEEE International Conference on Acoustics, Speech and Signal
  Processing}, 2007.

\bibitem{Book-LeBoudec}
J.-Y. {\noopsort{Boudec}}~Le~Boudec and P.~Thiran.
\newblock {\em Network Calculus}.
\newblock Springer Verlag, Lecture Notes in Computer Science, LNCS 2050, 2001.

\bibitem{BuLiCi11}
A.~Burchard, J.~Liebeherr, and F.~Ciucu.
\newblock On superlinear scaling of network delays.
\newblock {\em {IEEE/ACM} Transactions on Networking}, 19(4):1043--1056, Aug.
  2011.

\bibitem{choudhury96squeezing}
G.~Choudhury, D.~Lucantoni, and W.~Whitt.
\newblock Squeezing the most out of \text{ATM}.
\newblock {\em IEEE Transactions on Communications}, 44(2):203--217, Feb. 1996.

\bibitem{CiucuISIT11}
F.~Ciucu.
\newblock On the scaling of non-asymptotic capacity in multi-access networks
  with bursty traffic.
\newblock In {\em IEEE International Symposium on Information Theory (ISIT)},
  2011.

\bibitem{CiHoHu10}
F.~Ciucu, O.~Hohlfeld, and P.~Hui.
\newblock Non-asymptotic throughput and delay distributions in multi-hop
  wireless networks.
\newblock In {\em Allerton Conference on Communications, Control and
  Computing}, 2010.

\bibitem{CiPoSc13}
F.~Ciucu, F.~Poloczek, and J.~Schmitt.
\newblock Sharp bounds in stochastic network calculus.
\newblock {\em CoRR}, abs/1303.4114, 2013.

\bibitem{CiSc13}
F.~Ciucu and J.~Schmitt.
\newblock On the catalyzing effect of randomness on the per-flow throughput in
  wireless networks.
\newblock Technical report.
\newblock Available from
  \url{http://net.t-labs.tu-berlin.de/~florin/lib/randnet.pdf} and also from
  arXiv.org, 2013.

\bibitem{Ciucu12}
F.~Ciucu and J.~Schmitt.
\newblock Perspectives on network calculus - {No} free lunch but still good
  value.
\newblock In {\em ACM Sigcomm}, 2012.

\bibitem{Courcoubetis96}
C.~Courcoubetis and R.~Weber.
\newblock Buffer overflow asymptotics for a buffer handling many traffic
  sources.
\newblock {\em Journal of Applied Probability}, 33(3):886--903, Sept. 1996.

\bibitem{Couto03}
D.~S.~J. De~Couto, D.~Aguayo, B.~A. Chambers, and R.~Morris.
\newblock Performance of multihop wireless networks: shortest path is not
  enough.
\newblock {\em SIGCOMM Computer Communications Review}, 33(1):83--88, Jan.
  2003.

\bibitem{Durvy09}
M.~Durvy, O.~Dousse, and P.~Thiran.
\newblock Self-organization properties of {CSMA/CA} systems and their
  consequences on fairness.
\newblock {\em IEEE Transactions on Information Theory}, 55(3):931--943, Mar.
  2009.

\bibitem{Ephremides02}
A.~Ephremides.
\newblock Energy concerns in wireless networks.
\newblock {\em IEEE Wireless Communications}, 9(4):48--59, Aug. 2002.

\bibitem{EphremidesH98}
A.~Ephremides and B.~E. Hajek.
\newblock Information theory and communication networks: An unconsummated
  union.
\newblock {\em IEEE Transactions on Information Theory}, 44(6):2416--2434, Oct.
  1998.

\bibitem{FidlerFading06}
M.~Fidler.
\newblock A network calculus approach to probabilistic quality of service
  analysis of fading channels.
\newblock In {\em IEEE Globecom}, 2006.

\bibitem{Gallager85}
R.~G. Gallager.
\newblock A perspective on multiaccess channels.
\newblock {\em IEEE Transactions on Information Theory}, 31(2):124--142, Mar.
  1985.

\bibitem{GaoCL06}
Y.~Gao, D.-M. Chiu, and J.~C.~S. Lui.
\newblock Determining the end-to-end throughput capacity in multi-hop networks:
  methodology and applications.
\newblock In {\em ACM Sigmetrics/Performance}, pages 39--50, June 2006.

\bibitem{Gupta00}
P.~Gupta and P.~R. Kumar.
\newblock The capacity of wireless networks.
\newblock {\em IEEE Transactions on Information Theory}, 46(2):388--404, Mar.
  2000.

\bibitem{Haenggi05}
M.~Haenggi and D.~Puccinelli.
\newblock Routing in ad hoc networks: a case for long hops.
\newblock {\em IEEE Communications Magazine}, 43(10):93--101, Oct. 2005.

\bibitem{Heff8609:Markov}
H.~Heffes and D.~M. Lucantoni.
\newblock A {M}arkov modulated characterization of packetized voice and data
  traffic and related statistical multiplexer performance.
\newblock {\em IEEE Journal on Selected Areas in Communications},
  4(6):856--867, Sept. 1986.

\bibitem{jain03}
K.~Jain, J.~Padhye, V.~N. Padmanabhan, and L.~Qiu.
\newblock Impact of interference on multi-hop wireless network performance.
\newblock In {\em ACM Mobicom}, pages 66--80, 2003.

\bibitem{KleinrockSilv78}
L.~Kleinrock and J.~Silvester.
\newblock Optimum transmission radii for packet radio networks or why six is a
  magic number.
\newblock In {\em Proceedings of IEEE National Telecommunication Conference},
  pages 4.3.1--4.3.5, 1978.

\bibitem{LeeVaraiya03}
E.~A. Lee and P.~Varaiya.
\newblock {\em Structure and Interpretation of Signals and Systems}.
\newblock Addison-Wesley, 2003.

\bibitem{Li01}
J.~Li, C.~Blake, D.~S.~J. {De Couto}, H.~I. Lee, and R.~Morris.
\newblock Capacity of ad hoc wireless networks.
\newblock In {\em {ACM} Mobicom}, pages 61--69, 2001.

\bibitem{Mahmood11}
K.~Mahmood, M.~Vehkapera, and Y.~Jiang.
\newblock Delay constrained throughput analysis of {CDMA} using stochastic
  network calculus.
\newblock In {\em IEEE International Conference on Networks}, pages 83--88,
  2011.

\bibitem{Mergen05}
G.~Mergen and L.~Tong.
\newblock Stability and capacity of regular wireless networks.
\newblock {\em IEEE Transactions on Information Theory}, 51(6):1938--1953, June
  2005.

\bibitem{NeelyM05a}
M.~J. Neely and E.~Modiano.
\newblock Capacity and delay tradeoffs for ad hoc mobile networks.
\newblock {\em IEEE Transactions on Information Theory}, 51(6):1917--1937, June
  2005.

\bibitem{ShakkottaiLS10}
S.~Shakkottai, X.~Liu, and R.~Srikant.
\newblock The multicast capacity of large multihop wireless networks.
\newblock {\em IEEE/ACM Transactions on Networking}, 18(6):1691--1700, Dec.
  2010.

\bibitem{SharmaInf07}
G.~Sharma, N.~Shroff, and R.~Mazumdar.
\newblock Joint congestion control and distributed scheduling for throughput
  guarantees in wireless networks.
\newblock In {\em IEEE Infocom}, pages 2072--2080, 2007.

\bibitem{Silvester83}
J.~Silvester and L.~Kleinrock.
\newblock On the capacity of multihop slotted {ALOHA} networks with regular
  structure.
\newblock {\em IEEE Transactions on Communications}, 31(8):974--982, Aug. 1983.

\bibitem{Tang07}
J.~Tang and X.~Zhang.
\newblock Cross-layer modeling for quality of service guarantees over wireless
  links.
\newblock {\em IEEE Transactions on Wireless Communications}, 6(12):4504--4512,
  Dec. 2007.

\bibitem{WuNegi03}
D.~Wu and R.~Negi.
\newblock Effective capacity: {A} wireless link model for support of quality of
  service.
\newblock {\em IEEE Transactions on Wireless Communication}, 2(4):630--643,
  July 2003.

\bibitem{XieH09}
M.~Xie and M.~Haenggi.
\newblock Towards an end-to-end delay analysis of wireless multihop networks.
\newblock {\em Ad Hoc Networks}, 7(5):849--861, July 2009.

\bibitem{Zheng13}
K.~Zheng, F.~Liu, L.~Lei, C.~Lin, and Y.~Jiang.
\newblock Stochastic performance analysis of a wireless finite-state {Markov}
  channel.
\newblock {\em IEEE Transactions on Wireless Communications}, 12(2):782--793,
  Feb. 2013.

\bibitem{Zubaidy13}
H.~{\noopsort{Zubaidy}}Al-Zubaidy, J.~Liebeherr, and A.~Burchard.
\newblock A (min, $\times$) network calculus for multi-hop fading channels.
\newblock In {\em IEEE Infocom}, 2013.

\end{thebibliography}
%
%

\end{document}